\documentclass[12pt]{article}
\usepackage{amsmath,amsfonts,amssymb,mathtools,latexsym}
\usepackage[nosort]{cite}
\usepackage{hyperref}
\usepackage{graphicx}
\usepackage{color}
\usepackage{epsfig}
\usepackage{psfrag}	
\usepackage{tikz}
\usetikzlibrary{calc}
\usepackage{slashed}
\usepackage{bbold}
\usepackage{tensor}
\usepackage{ulem}
\usepackage[font=small,labelfont=bf,width=1\textwidth]{caption}
\usepackage{comment}
\usepackage{dsfont}
\usepackage[all]{xy}
\usepackage{multirow}
\usepackage{float}
\usepackage{pifont}
\usepackage{cite}
\usepackage{color}

\usepackage{titlesec} 
\usepackage{authblk} 

\usepackage{tocloft} 
\usepackage{ulem}
\usepackage{soul,xcolor}
\setstcolor{red}
\usepackage{appendix}
\usepackage{etoolbox}

\definecolor{ZKgreen}{rgb}{0,0.6,0.3}

\AtBeginEnvironment{subappendices}{%
  \addtocontents{toc}{\protect\addvspace{10pt}Appendices}
}
\makeatletter
\patchcmd{\@chapter}{\protect\numberline{\thechapter}#1}
{\@chapapp~\thechapter: #1}{}{}
\makeatother

\numberwithin{equation}{section}

\DeclareMathOperator{\tr}{tr}

\DeclareMathOperator{\Li}{Li}

\DeclareMathOperator{\vol}{vol}

\def\bF{{\mathbb{F}}}

\newcommand{\bea}{\begin{eqnarray}}
\newcommand{\eea}{\end{eqnarray}}
\newcommand{\beq}{\begin{equation}}
\newcommand{\eeq}{\end{equation}}
\newcommand{\bal}{\begin{equation}\begin{aligned}}
\newcommand{\eal}{\end{aligned} \end{equation}}

\newcommand{\cN}{{\mathcal N}}
\newcommand{\cP}{{\mathcal P}}

\newcommand{\cO}{{\mathcal O}}

\newcommand{\cW}{{\mathcal W}}
\newcommand{\cI}{{\mathcal I}}

\title{\begin{flushright}
\end{flushright}
\vspace{2 cm}
There and Back Again: \\ Bulk-to-Defect via Ward Identities
\vspace{5mm}}
\author{Jake Belton%
\thanks{\href{mailto:jake.belton@kcl.ac.uk}{jake.belton@kcl.ac.uk}}}
\author[2]{Ziwen Kong%
\thanks{\href{mailto:zwn.kong@gmail.com}{zwn.kong@gmail.com}}\normalsize}
\affil[1]{\it Department of Mathematics, King's College London,\protect\\London, 
WC2R 2LS, United Kingdom\vspace{4pt}}
\affil[2]{\it Deutsches Elektronen-Synchrotron DESY,
\protect\\
Notkestr.\ 85, 22607 Hamburg, Germany\vspace{4pt}}
\date{}

\begin{document}
\maketitle
\begin{abstract}
In conformal field theory, the presence of a defect may break the global symmetry, giving rise to defect operators such as the tilts. In this work, we derive integral identities that relate correlation functions involving bulk and defect operators---including tilts---to lower-point bulk--defect correlators, based on a detailed analysis of the Lie algebra of the symmetry group before and after the defect-induced symmetry breaking. As explicit examples, we illustrate these identities for the 1/2 BPS Maldacena--Wilson loop in $\cN=4$ SYM and for magnetic lines in the $O(N)$ model in $d=4-\varepsilon$ dimensions. We demonstrate that these identities provide a powerful tool both to check existing perturbative correlators and to impose nontrivial constraints on the CFT data.
\end{abstract}

\clearpage


\tableofcontents

\section{Introduction}
\label{sec:introduction}

Symmetry lies at the foundation of modern theoretical physics, serving as a guiding principle in the formulation and classification of quantum field theories. Among these, conformal field theories (CFTs) occupy a distinguished place, as they are defined and highly constrained by their underlying conformal symmetry. This symmetry extends the familiar Poincaré invariance to include scale transformations and special conformal transformations, leading to a rich algebraic structure that governs both the spectrum of local operators and the functional form of their correlation functions. In particular, the stringent constraints imposed by conformal invariance often render CFTs remarkably predictive, allowing many of their key properties to be determined from symmetry considerations alone.

Beyond the bulk theory, by which we mean the CFT in the absence of defects, many nontrivial conformal defects are known to exist. Such defects preserve a subalgebra of the bulk symmetry while breaking the remaining generators, providing tractable examples of symmetry breaking within a conformal framework. The introduction of conformal defects enriches the observable sector of a CFT, adding critical exponents and OPE coefficients intrinsic to the defect. The bulk CFT axioms
admit a straightforward extension that incorporates these defect observables and leads to
bootstrap relations connecting bulk and defect data \cite{Liendo:2012hy,Billo:2016cpy,Gadde:2016fbj}.
In this paper, we exploit classes of bulk--defect correlation functions to develop new and powerful constraints on both bulk and defect CFT data.

We consider a conformal defect $\cW$ of dimension $p$ in a $d$-dimensional CFT. Such a defect generally breaks translations in the transverse directions, which modifies the bulk stress-tensor Ward identity and gives rise to the displacement operator. This operator encodes the response of the defect to infinitesimal deformations of its position and plays a central role in characterizing its dynamics \cite{Billo:2016cpy}. Analogously, if the bulk theory possesses an internal symmetry $G$ that is broken to a subgroup $H$ by the defect, the corresponding Ward identity acquires a defect-localized modification:
\bal
\label{WardJandt}
\partial^{\mu} J_{\mu a} (x) = \delta^{d-p}(x_{\perp}) P_a^i \hat{t}_i(\tau)\,,
\eal
where $P_a^i$ is the projector that decomposes the bulk symmetry indices $a$ into the broken directions $i$ on the defect, and $x=(\tau,x_{\perp})$ where $\tau$ denotes coordinates along the defect and $x_\perp$ those transverse to the defect. The operators $\hat{t}_i$, referred to as tilt operators, inherit the defect’s conformal dimension 
$p$, making them exactly marginal. Exactly marginal operators are special in CFTs because they generate continuous deformations along the conformal manifold, a feature most commonly encountered in supersymmetric theories \cite{Leigh:1995ep, Green:2010da, Gomis:2015yaa, Giambrone:2021wsm}. Tilt operators, however, are more general: they naturally arise whenever a defect breaks an internal symmetry, and therefore appear in both supersymmetric and non-supersymmetric contexts \cite{Drukker:2022pxk, Herzog:2023dop, Sakkas:2024dvm, Belton:2025hbu, Bashmakov:2017rko}.

The normalization of the tilt operators is fixed by the normalization of the current $J_{\mu a}$ and locally determines the Zamolodchikov metric on the defect operator space
\bal
\label{Zmetric}
g_{ij} =\langle \hat{t}_i (0)\hat{t}_j(\infty)\rangle= C_t \delta_{ij}\,,
\eal
where $\hat{t} (\infty) =\lim\limits_{\tau\rightarrow \infty} |\tau|^{2p} \, \hat{t}(\tau)$. More generally, when a bulk operator $\cO$ is present, the Ward identity is modified accordingly to
\bal
\label{wardidentity}
\partial^{\mu} J_{\mu a} (x) \cO (x') =\delta_{\cW}^{d-p} (x_{\perp}) P_a^i \hat{t}_i (\tau) \cO (x') +\delta^d (x-x') (T_a \cO) (x)\,.
\eal
where $T_a$ represents the action of the symmetry generator corresponding to the current 
$J_{\mu a}$ on the bulk operator $\cO$. Allowing $\cO$ in a general representation of the symmetry group $G$, we label the corresponding representation space by indices $\alpha$, and encode them through a bulk coupling $r^{\alpha}$. 

In this way, one may define a unified generating functional that simultaneously captures the dynamics of both bulk and tilt operators,
\bal
\label{principle}
Z[r,w]=\int D\phi\, e^{-S} e^{\int d^d x\, r^{\alpha} (x) \cO_{\alpha} (x)} \cW[e^{\int d^p \tau\, w^i (\tau) \hat{t}_i (\tau)}]\,.
\eal
Here $\cW[\cdots]$ denotes insertions into the defect $\cW$, so the expression
$\cW[e^{\int d^p \tau\, w^i (\tau) \hat{t}_i (\tau)}]$ serves as a generating
functional for correlators of tilt operators $\hat{t}_{i}$ on the defect worldvolume. The functions $w_i(\tau)$ parametrize the directions in the symmetry algebra that are broken by the presence of the defect and take values in $\mathfrak{h}^{\perp}\equiv \mathfrak{g}/\mathfrak{h}$. In particular, connected correlators of tilt operators localized on the defect can be obtained by functional differentiation with respect to $w$,
\bal
\langle \hat{t}_{i_1} (\tau_1) \cdots \hat{t}_{i_n} (\tau_n) \rangle_c =\frac{\delta}{\delta w^{i_1}(\tau_1)} \cdots \frac{\delta}{\delta w^{i_n}(\tau_n)} \log Z[r,w]\Big|_{r=w=0}\,,
\eal
and more generally, one can evaluate correlators involving both bulk and defect operators,
\begin{multline}
\label{connectedbulkdefect}
\langle \cO_{\alpha_1} (x_1) \cdots \cO_{\alpha_m} (x_m) \hat{t}_{i_1} (\tau_1) \cdots \hat{t}_{i_n} (\tau_n) \rangle_c \\
=\frac{\delta}{\delta r^{\alpha_1} (x_1)} \cdots \frac{\delta}{\delta r^{\alpha_m} (x_m)} \frac{\delta}{\delta w^{i_1}(\tau_1)} \cdots \frac{\delta}{\delta w^{i_n}(\tau_n)} \log Z[r,w]\Big|_{r=w=0}\,.
\end{multline}

In this work, we establish integral identities that apply to arbitrary numbers of bulk
operators and any number of tilts, relating integrated bulk–defect correlators with multiple
tilt insertions to corresponding lower-point, non-integrated correlators. These relations
impose nontrivial constraints on the CFT data: they can be used to verify the consistency of
existing results in the literature or, conversely, to extract new CFT data that could be
difficult to obtain through more traditional methods such as localization or explicit
diagrammatic computations. In addition, we identify which classes of these identities remain invariant under changes of renormalization scheme and which do not, and we show how suitable constructions remove all scheme-dependent ambiguities.

The rest of the paper is organized as follows. Section~\ref{sec:derivation}
presents the derivation of the integral identities, discusses potential scheme dependence
and contact terms. In Sections~\ref{sec:N=4SYM} and \ref{sec:O(N)}, we illustrate the application of these integral identities in explicit examples, including 1/2 BPS Wilson loops in $\cN=4$ super Yang--Mills and magnetic lines in the $O(N)$ model, and demonstrate how they constrain the relevant CFT data and produce new predictions. In Section~\ref{sec:O(N)}, we further compute several new correlators that are incorporated into the integral identities, particularly those involving tilts and unprotected operators.

\textbf{Note}: During the completion of this paper, independent work~\cite{Girault:2025kzt} appeared last week, with results that partially overlap with ours. There is another paper in progress~\cite{upcoming} on which we heavily rely, whose conventions and techniques we adopt throughout this work.

\section{Derivation}
\label{sec:derivation}
We now present the general framework in which we derive identities between connected correlators. Many of the foundations that follow are based on work in \cite{upcoming}, where a more detailed discussion can be found.

Let $L_g$ denote the left action of $g \in G$ on the bulk and defect couplings $r$ and $w$.
Under such a transformation the partition function behaves as
\bal \label{Partition fn transformation}
Z[L_g(r),L_g(w)]=e^{A[g,w]} Z[r, w]\,,
\eal
where $A[g,w]$ is a local functional of $w$, and $e^{A[g,w]}$ encodes a possible anomaly
generated by the action of the symmetry group $G$ on the partition function. As emphasized in \cite{upcoming}, anomalies play an important role, including in bulk--defect systems. However, in the examples studied in the following sections no anomalies arise, so for the remainder of this work we restrict to the case $e^{A[g,w]} = 1$ and leave the analysis of nontrivial anomalies to future work.

For a more systematic analysis, we expand around the identity of the group by taking 
$g=e^{\lambda}$, with $\lambda$ infinitesimal. Then the left action $L_g$ can be linearized as
\bal
L_{e^{\lambda}}(r)=r+\rho(\lambda, r) +O(\lambda^2)\,,\qquad L_{e^{\lambda}}(w) =w+l(\lambda,w) +O(\lambda^2)\,,
\eal
where $l(\lambda,w)$ and $\rho(\lambda,r)$ are vector fields linear in $\lambda$ that generate infinitesimal transformations of $w$ and $r$ respectively under the Lie algebra $\mathfrak{g}$.

First, the infinitesimal transformation of the bulk coupling $r^{\alpha}$ can be expressed as
\begin{equation}
    \rho(\lambda, r) = \lambda^{i}\, r^{\alpha}\,[T_{i}^{(R)} ]_{\alpha},
\end{equation}
where $T_{i}^{(R)}$ denotes the action of the generator $e_{i}$ on the basis 
$e_{\alpha}^{(R)}$ of the representation $R$ in which the operator $\cO$ transforms.
For notational simplicity we drop the label $(R)$ in the following.

The infinitesimal transformation of $w$ generated by $l(\lambda,w)$ is more subtle. When $\lambda$ is an element of the unbroken subalgebra $\mathfrak{h}$, the transformation reduces to the familiar adjoint action,
\bal
l(\lambda,w)=[\lambda,w]\,.
\eal
For a general $\lambda \in \mathfrak{g}$, nonlinear effects appear, which can be organized as a formal power series in $w$,
\bal
l(\lambda,w)=\sum_{n=0}^{\infty} \frac{1}{n!} l_n(\lambda;w,\cdots,w)\,,
\eal
with the leading term $l_0(\lambda)=\lambda^{\perp} \in \mathfrak{h}^{\perp}$ capturing the projection of $\lambda$ onto the broken directions, and higher-order terms $l_n$ ($n\ge 1$) encoding nonlinear and renormalization scheme dependent contributions. Importantly, $l(\lambda,w)$ must respect the algebra $\mathfrak{g}$, which is expressed via the commutator relation
\bal
\label{lambdaalgebra}
\left[l^i (\lambda_1,w) \frac{\partial}{\partial w^i}, l^j (\lambda_2,w) \frac{\partial}{\partial w^j}\right] =-l^i ([\lambda_1,\lambda_2],w) \frac{\partial}{\partial w^i}\,. 
\eal
Focusing on the broken directions, $\lambda \in \mathfrak{h}^{\perp}$, the commutator can be expanded order by order in $w$. Setting $\lambda_1=\lambda$, $\lambda_2=w$, one finds a recursive structure among the $l_n$,
\begin{multline}
l_{n+1}(\lambda;w,w,\cdots,w)- l_{n+1}(w;\lambda,w,\cdots,w)= l_n([\lambda,w];w,\cdots,w)\\
-\sum_{k=1}^n \begin{pmatrix}
    n \\k
\end{pmatrix} (l_{n-k+1} (\lambda;l_k(w;w,\cdots,w),w,\cdots,w)-l_{n-k+1} (w;l_k(\lambda;w,\cdots,w),w,\cdots,w))\,.
\end{multline}
The first two explicit orders read:
\bal
\label{initiallambdaalgebra}
l_1(\lambda;w) -l_1(w;\lambda) &=l_0([\lambda,w]) =[\lambda,w]^{\perp}\,,\\
l_2(\lambda;w,w) -l_2(w;\lambda,w) &= l_1([\lambda,w];w)-l_1(\lambda;l_1(w;w)) +l_1(w;l_1(\lambda;w))\,.
\eal


With all these ingredients in place, we can take the logarithm of \eqref{Partition fn transformation} and, expanding to order $r^m$ and $w^n$, one obtains identities among the connected correlators,
\begin{multline}
\label{generalintid}
\sum_{i=1}^m \langle \cO(r_1) \cdots \cO(\rho(\lambda,r_i)) \cdots \cO(r_m) \hat{t}(w)^n\rangle_c \\
+ \sum_{k=0}^n \begin{pmatrix}
    n\\k
\end{pmatrix}\langle \cO(r_1) \cdots \cO(r_m) \hat{t}(l_k(\lambda;w,\cdots,w))\hat{t}^{n-k} (w) \rangle_c =0 \,.
\end{multline}
where we define $\cO(r)\equiv \int d^d x\, r^{\alpha} (x) \cO_{\alpha}(x)$, and $\hat{t}(w)\equiv \int d^p \tau\, w^i (\tau) \hat{t}_i(\tau)$. 
From now on we focus on the case where $\lambda\in \mathfrak{h}^{\perp}$.

\subsection{\texorpdfstring{$\boldsymbol{m=0}$}{m=0}}
\label{sec:m=0}
When $m=0$, we return to the defect-only case. A much more detailed treatment, including the anomaly terms and all associated subtleties, as well as the renormalization scheme dependence, is provided in \cite{upcoming}. However, to make the discussion more complete and to facilitate the use of these results in the $m\ne 0$ cases later, we provide only a brief review and list the material needed subsequently.

For an $n$-point function, having all points on the defect leads to fewer cross-ratios compared to the general setup with $m$ bulk and 
$n-m$ defect operators. This simplification provides a more tractable way to fix the $l_n$. In this way, we recover the integral identities in \cite{upcoming}
\bal
\sum_{k=0}^n \begin{pmatrix}
    n\\k
\end{pmatrix} \langle \hat{t}(l_k(\lambda;w,\cdots,w))\hat{t}^{n-k} (w) \rangle_c =0 \,.
\eal

\subsubsection{\texorpdfstring{$\boldsymbol{n=0}$}{n=0}}
\bal
\label{m=0n=0}
\langle \hat{t}(\lambda) \rangle_c =0\,.
\eal
After acting with $\frac{\partial}{\partial \lambda^i}$, the equation reads
\bal
\label{m=0n=0der}
\int d^p\tau_1 \langle \hat{t}_i(\tau_1) \rangle_c =0\,.
\eal
In the case of flat defects the condition is automatically satisfied, as conformal invariance implies the vanishing of the one-point function, $\langle \hat{t} \rangle=0$. From now on, we impose this and omit defect one-point functions in the subsequent expressions.

\subsubsection{\texorpdfstring{$\boldsymbol{n=1}$}{n=1}}
\bal
\langle \hat{t}(\lambda) \hat{t}(w) \rangle_c =0\,.
\eal
Applying $\frac{\partial}{\partial \lambda^i} \frac{\delta}{\delta w^j (\tau_2)}$ to the above expression yields
\bal
\int d^p \tau_1 \langle \hat{t}_i(\tau_1) \hat{t}_j(\tau_2) \rangle_c =0\,.
\eal
At non-coincident points, the two-point function reads
\bal
\langle \hat{t}_i(\tau_1) \hat{t}_j(\tau_2) \rangle_c=\frac{g_{ij}}{|\tau_1 -\tau_2|^{2p}}\,,
\eal
where $g_{ij}$ is the  Zamolodchikov metric defined in \eqref{Zmetric}, plus the distributional extension of the two-point function at coincident points $\tau_1=\tau_2$. A more detailed analysis in \cite{upcoming} demonstrates that any allowed distribution is consistent with the scale invariance of the integral, ensuring that the condition is satisfied.

\subsubsection{\texorpdfstring{$\boldsymbol{n=2}$}{n=2}}
\bal
\langle \hat{t}(\lambda) \hat{t}(w)^2 \rangle_c +2\langle \hat{t}(l_1(\lambda;w)) \hat{t}(w) \rangle_c  =0\,.
\eal
At this order, for the first time the expression contains a non-vanishing, scheme-dependent term $l_1$. Taking the derivative $\frac{\partial}{\partial \lambda^i} \frac{\delta}{\delta w^j(\tau_2)} \frac{\delta}{\delta w^k(\tau_3)}$ of the above expression results in
\begin{multline}
\label{m=0n=2}
\int d^p \tau_1 \langle \hat{t}_i(\tau_1) \hat{t}_j(\tau_2) \hat{t}_k (\tau_3) \rangle_c +l_1^s (e_i;e_j) \langle \hat{t}_s (\tau_2) \hat{t}_k(\tau_3) \rangle_c +l_1^s(e_i;e_k) \langle \hat{t}_s(\tau_3) \hat{t}_j(\tau_2) \rangle_c  
=0\,.
\end{multline}
Generally, for arbitrary dimension $p$, the three-point function takes the form\footnote{An exception is the oriented line defect with $p=1$, where the three-point function is of the form
\bal
\langle \hat{t}_i (\tau_1) \hat{t}_j (\tau_2) \hat{t}_k (\tau_3) \rangle=\frac{f_{ijk}}{(\tau_1 -\tau_2) (\tau_2 -\tau_3) (\tau_3 -\tau_1)}\,,
\eal
with $f_{ijk}$ totally antisymmetric. This produces a non-vanishing anomaly term; nevertheless, as discussed in \cite{upcoming}, the integral of such a correlator vanishes.}
\bal
\langle \hat{t}_i (\tau_1) \hat{t}_j (\tau_2) \hat{t}_k (\tau_3) \rangle =\frac{d_{ijk}}{|\tau_1 -\tau_2|^p |\tau_2 -\tau_3|^p |\tau_3 -\tau_1|^p}\,.
\eal
However, one can show that $d_{ijk}$ must vanish identically. This has a natural interpretation from the perspective of the defect conformal manifold: the three-point function of exactly marginal operators must vanish, since a nonzero value would generate a nonvanishing beta function, thereby violating the defining condition of conformality on the manifold. It follows that the integrated tilt three-point function vanishes identically, including at coincident points $\tau_2=\tau_3$. Consequently, \eqref{m=0n=2} transforms into
\bal
l_1^s (e_i;e_j) \langle \hat{t}_s (\tau_2) \hat{t}_k(\tau_3) \rangle_c +l_1^s(e_i;e_k) \langle \hat{t}_s(\tau_3) \hat{t}_j(\tau_2) \rangle_c =0\,,
\eal
combined with \eqref{initiallambdaalgebra}, this allows one to solve for $l_1$ explicitly,
\bal
l_1^s (e_j,e_k) =\frac{1}{2} \bigl( [e_j,e_k]^s + [e^s,e_k]_j+ [e^s,e_j]_k \bigl)\,,
\eal
where we use $\delta_{ij}$ and $\delta^{ij}$ to raise and lower indices. In particular, if $[\mathfrak{h}^{\perp},\mathfrak{h}^{\perp}]\subseteq \mathfrak{h}$---which occurs for the magnetic line in the $O(N)$ model and for 1/2 BPS Wilson loops in $\cN=4$ super Yang-Mills---the above expression shows that $l_1(\lambda;w)=0$. We will use this result to simplify most of the expressions that follow.

\subsubsection{\texorpdfstring{$\boldsymbol{n=3}$}{n=3}}
\label{sec:m=0n=3}
\bal
\langle \hat{t}(\lambda) \hat{t}(w)^3 \rangle_c +3\langle \hat{t}(l_1(\lambda;w)) \hat{t}(w)^2 \rangle_c +3\langle \hat{t}(l_2(\lambda;w,w)) \hat{t}(w) \rangle_c =0\,.
\eal
We restrict to $[\mathfrak{h}^{\perp},\mathfrak{h}^{\perp}]\subseteq \mathfrak{h}$, for which $l_1(\lambda;w)=0$. There are, however, situations where $l_1(\lambda;w) \neq 0$; see, for example,~\cite{Drukker:2022txy, Bliard:2024bcz}. Differentiating with $\frac{\partial}{\partial \lambda^i} \frac{\delta}{\delta w^j (\tau_2)} \frac{\delta}{\delta w^k (\tau_3)} \frac{\delta}{\delta w^l (\tau_4)}$ results in
\begin{multline}\label{m=0 n=3 identity}
\int d^p \tau_1 \langle \hat{t}_i (\tau_1) \hat{t}_j (\tau_2) \hat{t}_k (\tau_3) \hat{t}_m (\tau_4) \rangle_c +\delta^p (\tau_2 -\tau_3) l_2^s (e_i;e_j,e_k) \langle \hat{t}_s(\tau_2) \hat{t}_m (\tau_4) \rangle \\
+\delta^p (\tau_2 -\tau_4) l_2^s (e_i;e_j,e_m) \langle \hat{t}_s (\tau_2) \hat{t}_k(\tau_3) \rangle +\delta^p (\tau_3 -\tau_4) l_2^s (e_i;e_k,e_m) \langle \hat{t}_s(\tau_3) \hat{t}_j (\tau_2) \rangle = 0\,.
\end{multline}

The first identity can be extracted by considering distinct insertion points. For simplicity, we set $\tau_2 = 0$, $\tau_3 = 1$, and $\tau_4 = \infty$, in which case all $\delta$-function terms vanish and the above expression reduces to
\bal
\label{nolog}
\int d^p \tau_1 \langle \hat{t}_i (\tau_1) \hat{t}_j (0) \hat{t}_k (1) \hat{t}_m (\infty) \rangle_c=0\,.
\eal
We should be cautious about potential divergences as $\tau_1$ approaches $\tau_2$, $\tau_3$, or $\tau_4$. Since this only-defect issue is not central to the present discussion, we refer the reader to \cite{upcoming} for a detailed treatment. Related considerations will, however, appear again in Section~\ref{sec:m=1n=1}.

We can obtain more information from \eqref{m=0 n=3 identity} by further integrating with respect to $\tau_2$ while keeping $\tau_3\neq\tau_4$. Let $\varphi(\tau_2)$ be an arbitrary measure satisfying $\varphi(\tau_4)=0$. Upon integrating the above expression once more against this measure we obtain
\bal
\label{doubleintegraldefect}
\int d^p \tau_1 d^p \tau_2 \varphi(\tau_2) \langle \hat{t}_i (\tau_1) \hat{t}_j (\tau_2) \hat{t}_k (\tau_3) \hat{t}_m (\tau_4) \rangle_c +\varphi(\tau_3) l_2^s (e_i;e_j,e_k) \langle \hat{t}_s(\tau_3) \hat{t}_m (\tau_4) \rangle =0\,.
\eal
A convenient choice is to place $\tau_4$ at infinity, $\tau_3$ at zero, and normalize $\varphi(\tau_3)=1$. Although this double integration is scheme-dependent---since it corresponds to the $l_2$ term---it can, for the same reason, be employed to determine $l_2$, where
\bal
\label{l2}
l_2^s (e_i;e_j,e_k)=-\int d^p \tau_1 d^p \tau_2 \varphi(\tau_2) \langle \hat{t}_i (\tau_1) \hat{t}_j (\tau_2) \hat{t}_k (0) \hat{t}^s (\infty) \rangle_c\,,
\eal
where we define $\hat{t}^s(\infty)=\frac{\delta^{rs}}{C_t} \hat{t}_r(\infty)$. In this case, we encounter not only potential divergences when two tilts coincide, but also additional divergences when three tilts collide, i.e. both $\tau_2 ,\tau_3 \to \tau_1$. These may lead to logarithmic divergences, necessitating regularization within a suitable renormalization scheme. The detailed discussion is deferred to Section~\ref{sec:m=1n=2}, or see \cite{upcoming} for a more comprehensive treatment. However, by taking a commutator and using \eqref{initiallambdaalgebra},
\bal
l_2(e_i;e_j,e_k)- l_2(e_j;e_i,e_k) =[[e_i,e_j],e_k]\,,
\eal
we find that the scheme-dependent part cancels out, leading to a scheme-independent identity
\bal
\label{logid}
\int d^p \tau_1 d^p \tau_2 (\varphi(\tau_2) -\varphi(\tau_1)) \langle \hat{t}_i (\tau_1) \hat{t}_j (\tau_2) \hat{t}_k (0) \hat{t}_m (\infty) \rangle_c + [[e_i,e_j],e_k]^s \langle \hat{t}_s(0) \hat{t}_m (\infty) \rangle =0\,.
\eal
Since the four-point function depends only on the cross-ratio, one of the two integrations can be performed explicitly. Consequently, in any renormalization scheme, the expression above can be reduced to a $\varphi$-independent\footnote{The result actually depends only on the value of $\varphi(0)$, which we have normalized to be 1.} form,
\bal
\label{logexpression}
\vol_{S^{p-1}} \int d^p\tau \log|\tau| \langle \hat{t}_{i}(1) \hat{t}_{j}(\tau) \hat{t}_k (0) \hat{t}_l(\infty)\rangle_c=[[e_i,e_j],e_k]^s \langle \hat{t}_s(\tau_3) \hat{t}_m (\tau_4) \rangle\,.
\eal
This result coincides with the integral identities relating the integrated four-tilt correlators to the curvature tensor of the defect conformal manifold, as discussed in \cite{Drukker:2022pxk, upcoming, Kong:2025sbk, Friedan:2012hi}.


\subsection{\texorpdfstring{$\boldsymbol{m=1}$}{m=1}}
\label{sec:m=1}
We next consider the situation where bulk operators are included. Note that, compared to the defect-only case in section \ref{sec:m=0}, $l_n$ now appears in terms at $O(w^n)$ rather than $O(w^{n+1})$, i.e. it is shifted down by one. The reason is that in the defect-only case, $l_n$ would only appear in $\langle \hat{t}(l_n(\lambda;w,\cdots,w)\rangle_c$ but this term vanishes since it is a defect one-point function. In the presence of bulk operators, this term is replaced by $\langle \cO_1(r_1) \cdots O_m(r_m) \hat{t}(l_n(\lambda;w,\cdots,w)) \rangle_c$, which is non-zero in general. The simplest scenario involves a single bulk operator, in which case \eqref{generalintid} becomes
\bal
\label{m=1intid}
\langle \cO(\rho(\lambda,r)) \hat{t}(w)^n\rangle_c + \sum_{k=0}^n \begin{pmatrix}
    n\\k
\end{pmatrix}\langle \cO(r) \hat{t}(l_k(\lambda;w,\cdots,w))\hat{t}^{n-k} (w) \rangle_c =0 \,.
\eal

\subsubsection{\texorpdfstring{$\boldsymbol{n=0}$}{n=0}}
\bal
\langle \cO(\rho(\lambda,r))\rangle_c + \langle \cO(r) \hat{t}(\lambda) \rangle_c =0\,.
\eal
This constitutes an identity between the bulk one-point function and the integral over the defect of the bulk–defect two-point function. Applying $\frac{\partial}{\partial \lambda^i} \frac{\delta}{\delta r^{\alpha}(x_0)}$ to the above, we obtain
\bal
\label{m=1n=0der}
(T_i)_{\alpha}^{\beta} \langle \cO_{\beta}(x_0)\rangle_c + \int d^p\tau_1 \langle \cO_{\alpha} (x_0) \hat{t}_i(\tau_1) \rangle_c =0\,.
\eal
This identity provides an explicit relation between the CFT data. 
To see this, recall that the bulk and bulk--defect correlation functions that appear in \eqref{m=1n=0der} take the form \cite{Billo:2016cpy}
\bal\label{bulk1pt and bulk-defect2pt for general operators}
\langle \cO_\alpha(x_0)\rangle &= \mathcal{T}_\alpha \frac{a_{\cO }}{|x_{0\perp}|^{\Delta_{\cO}}}\,,\\
\langle \cO_\alpha(x_0) \hat{t}_i(\tau_1) \rangle &= \mathcal{T}'_{\alpha ; i} \frac{b_{\cO \hat{t}}}{|x_{0\perp}|^{\Delta_{\cO} -p} |x_{01}|^{2p} }\,,
\eal
where $\mathcal{T}_\alpha$ and $\mathcal{T}'_{\alpha ; i}$ are some tensor structures dependent on the defect and the representation in which $\mathcal{O}_\alpha$ transforms.
The integral over the functional part of the bulk--defect two-point function can be done explicitly, with the result relying on
\bal
\int \frac{d^{p}\tau_1}{|x_{01}|^{2p}} = \vol_{S^{p-1}} \int \frac{dr r^{p-1}}{(|x_{0\perp}|^2+ r^2)^{p} } = \frac{2^{1-p} \pi^{\frac{p+1}{2}}}{\Gamma\left(\frac{p+1}{2} \right) |x_{0\perp}|^p}\,,
\eal
where
\bal
\label{volS}
\vol_{S^{p-1}} = \frac{2 \pi^{p/2}}{\Gamma\left(\frac{p}{2} \right)}
\eal
is the volume of a $(p-1)$-dimensional unit sphere. Using this in \eqref{m=1n=0der}, we find
\begin{equation}\label{m=1n=0General}
    (T_i)_\alpha^\beta \mathcal{T}_\beta a_{\mathcal{O}} + \mathcal{T}'_{\alpha ; i} b_{\mathcal{O}\hat{t}} \frac{2^{1-p} \pi^{\frac{p+1}{2}}}{\Gamma\left(\frac{p+1}{2} \right)} = 0\,.
\end{equation}

To make this relation concrete, let us consider a bulk theory of $N$ scalar fields $\phi_I$
($I = 1,\ldots,N$) endowed with an $O(N)$ global symmetry, and introduce a rank-$J$
operator $\cO_J$ transforming in the symmetric traceless representation
\begin{align}
\label{cOJ}
\cO_J(u)
= u^{I_1}\cdots u^{I_J}\, \phi_{I_1}\cdots \phi_{I_J}
= (u\cdot \phi)^J\,,
\end{align}
where $u^I$ is an $N$-component null polarization vector $u^I$ satisfying $u^2 = 0$. 
Suppose our defect couples to a polarization vector $\theta$ that induces the symmetry breaking $O(N)\rightarrow O(N-1)$---we can choose $\theta^i = \delta^{iN}$---as happens in the 1/2 BPS Wilson line in $\mathcal{N}=4$ super Yang--Mills\footnote{We slightly abuse the use of the symbol $N$. In the 1/2 BPS Wilson line, the symmetry breaking $O(N)\rightarrow O(N-1)$ occurs for $N=6$---this is the breaking of the R-symmetry induced by the presence of the Wilson line. In light of this, in Section \ref{sec:N=4SYM} the symbol $N$ then takes on the role of the rank of the gauge group $SU(N)$.}, discussed in \ref{sec:N=4SYM} and in the magnetic line defect in the $O(N)$ model, discussed in \ref{sec:O(N)}.
We choose the tensor structures $\mathcal{T}_{I_1 \cdots I_J}$ and $\mathcal{T}'_{I_1 \cdots I_J ; i}$ to be
\begin{align}\label{tensors}
    \mathcal{T}_{I_1 \cdots I_J} & = \delta_{I_1 N} \cdots \delta_{I_J N}\,,\\
    \mathcal{T}'_{I_1 \cdots I_J ; i} & = \sum_{s=1}^J \delta_{I_1 N} \cdots \delta_{I_s i} \cdots \delta_{I_J N}\,.
\end{align}
Moreover, the charges $T_i$ act on each constituent $\phi_{I_s}$ of $\mathcal{O}_J$ in turn and we find
\begin{equation}\label{OJ transformation}
    \mathcal{O}_J (u) \rightarrow u^{I_1}\cdots u^{I_J}  \sum_{s=1}^J (T_i)_{I_s}^I \phi_{I_1}\cdots \phi_I \cdots \phi_{I_J} = J(u\cdot\phi)^{J-1}((u\cdot\theta)\phi_i - (\theta\cdot\phi)u_i)\,.
\end{equation}
When considering the bulk one-point function of the operator above, only the second term survives and the coefficient of $a_\mathcal{O}$ in \eqref{m=1n=0General} becomes $-J(u\cdot\theta)^{J-1}(u\cdot\hat{u})$, where we have also contracted with a null polarization vector $\hat{u}$ which is in the preserved symmetry group $O(N-1)$ and satisfies $\hat{u}\cdot\theta = \hat{u}^2=0$. 
Contracting $\mathcal{T}'_{I_1 \cdots I_J ; i}$ with the same polarization vectors gives $J(u\cdot\theta)^{J-1}(u\cdot\hat{u})$.
After these manipulations, \eqref{m=1n=0General} takes the form
\bal
\label{tiltwardid}
a_{\cO_J} = \frac{2^{1-p} \pi^{\frac{p+1}{2}}}{\Gamma\left(\frac{p+1}{2} \right)} b_{\cO_J t}\,.
\eal
This identity was originally derived in \cite{Padayasi:2021sik}. Moreover, the analogous identity for the displacement operator has already been found in \cite{Billo:2016cpy}.
For $p=1$, which will be relevant for examples covered later, \eqref{tiltwardid} becomes
\bal
\label{tiltwardid1d}
a_{\cO_J} =\pi b_{\cO_J t}\,.
\eal

\subsubsection{\texorpdfstring{$\boldsymbol{n=1}$}{n=1}}
\label{sec:m=1n=1}
\bal
\langle \cO(\rho(\lambda,r)) \hat{t}(w)\rangle_c + \langle \cO(r) \hat{t}(\lambda)\hat{t}(w) \rangle_c + \langle \cO(r) \hat{t}(l_1(\lambda;w)) \rangle_c =0 \,.
\eal
Acting with $\frac{\partial}{\partial \lambda^i} \frac{\delta}{\delta w^j (\tau_2)}  \frac{\delta}{\delta r^{\alpha}(x_0)}$, the above expression becomes
\bal
(T_i)_{\alpha}^{\beta} \langle \cO_{\beta}(x_0) \hat{t}_j(\tau_2)\rangle_c + \int d^p\tau_1 \langle \cO_{\alpha}( x_0) \hat{t}_i(\tau_1) \hat{t}_j(\tau_2) \rangle_c + l_1^s (e_i;e_j)\langle \cO_{\alpha}(x_0) \hat{t}_s(\tau_2) \rangle_c =0 \,.
\eal
In the case $[\mathfrak{h}^{\perp},\mathfrak{h}^{\perp}]\subseteq \mathfrak{h}$ for which $l_1 (e_i,e_j)=0$, this reduces to
\bal
\label{m=1n=1singleint}
(T_i)_{\alpha}^{\beta} \langle \cO_{\beta}(x_0) \hat{t}_j(\tau_2)\rangle_c + \int d^p\tau_1 \langle \cO_{\alpha}( x_0) \hat{t}_i(\tau_1) \hat{t}_j(\tau_2) \rangle_c =0 \,.
\eal
Returning to the bulk operator $\cO_J$ introduced in \eqref{cOJ} for $J\ge 2$, its three-point function with two tilts is given by
\begin{multline}
\label{uOtt}
\langle \cO_J(u, x_0) (\hat{u}_1 \cdot \hat{t})(\tau_1) (\hat{u}_2 \cdot \hat{t})(\tau_2) \rangle_c \\
= \frac{(u\cdot \theta)^{J-2}}{|x_{0\perp}|^{\Delta_{\cO_J}-2p} |x_{01}|^{2p} |x_{02}|^{2p}} \left((u\cdot \hat{u}_1) (u\cdot \hat{u}_2) F_{1c}(\chi) +(u\cdot \theta)^2 (\hat{u}_1 \cdot \hat{u}_2) \frac{1}{\chi^p}F_{2c}(\chi) \right)\,,
\end{multline}
where $F_{1c}(\chi)$ and $F_{2c}(\chi)$ denote the functional dependence on the
conformal cross-ratio
\begin{align}
\label{chi cross ratio}
\chi = \frac{|x_{0\perp}|^{2}\, |\tau_{12}|^{2}}
            {|x_{01}|^{2}\, |x_{02}|^{2}}\,,
\end{align}
which appears in connected bulk--tilt--tilt correlators. Here $|\tau_{ij}|^2 = (\tau_i - \tau_j)^2$ and $|x_{ij}|^2 = (\tau_i - \tau_j)^2 + (x_{i\perp} - x_{j\perp})^2$.
Upon contracting the indices in \eqref{m=1n=1singleint} with the corresponding polarization vectors $u$ and $\hat{u}_{1,2}$, it gives
\begin{multline}
\int d^p \tau_1 \langle \cO_J(u, x_0) (\hat{u}_1 \cdot \hat{t})(\tau_1) (\hat{u}_2 \cdot \hat{t})(\tau_2) \rangle_c \\
= \left( J(J-1)(u\cdot \theta)^{J-2} (u \cdot \hat{u}_1)(u \cdot \hat{u}_2) -J (u\cdot \theta)^{J} (\hat{u}_1 \cdot \hat{u}_2) \right) \frac{b_{\cO_J \hat{t}}}{|x_{0\perp}|^{\Delta_{\cO_J} -p} |x_{02}|^{2p}}\,,
\end{multline}
Inserting \eqref{uOtt} into the above expression and carrying out the integration, comparison of the tensor structures leads to two independent identities for $F_1$ and $F_2$,
\begin{align}
\label{intF1}
\frac{1}{2} \vol_{S^{p-1}} \int_0^1 \frac{d\chi}{(\chi(1-\chi))^{1-p/2}} F_{1c}(\chi) &=J(J-1) b_{\cO_J \hat{t}}\,,\\
\label{intF2}
\frac{1}{2} \vol_{S^{p-1}} \int_0^1 \frac{d\chi}{\chi^{1+p/2} (1-\chi)^{1-p/2}} F_{2c}(\chi) &=-J b_{\cO_J \hat{t}}\,,
\end{align}
where $\vol_{S^{p-1}}$ is given in \eqref{volS}.

Before proceeding, let us briefly discuss the two identities derived above. Since the correlators in \eqref{m=1n=1singleint} are taken to be their connected parts, whereas in much of the literature the correlators are presented in their full (including disconnected) form, we shall clarify how to extract the connected part from them. We note that for $\langle \cO \hat{t} \rangle$, the disconnected contribution vanishes, so $\langle \cO \hat{t} \rangle_c = \langle \cO \hat{t} \rangle$. In contrast, for $\langle \cO \hat{t} \hat{t} \rangle$, the connected component is
\bal
\label{disconnectedOtt}
\langle \cO \hat{t}\hat{t}\rangle_c=\langle \cO \hat{t}\hat{t}\rangle -\langle \cO \rangle \langle \hat{t}\hat{t}\rangle\,,
\eal
implying that
\bal
\label{disconnectedF}
F_{1c}=F_1,\qquad\qquad F_{2c}=F_2 - a_{\cO_J} C_{\hat{t}}\,.
\eal

Another subtlety arises in perturbative CFTs. If there exists an operator $\hat{\phi}$ in the $\theta$ direction appearing in the $\hat{t} \times \hat{t}$ OPE with scaling dimension $\Delta_{\hat{\phi}} = p + \gamma_{\hat{\phi}}$, where $p$ is the classical dimension and
$\gamma_{\hat{\phi}}$ is its anomalous dimension. Then, as $\tau_{1}\rightarrow \tau_{2}$,
\bal
\hat{t}_i (\tau_1) \times \hat{t}_j (\tau_2) =\delta_{ij} C_{\hat{t}} \mathbb{1} |\tau_{12}|^{-2p}+ \delta_{ij} \lambda_{\hat{t}\hat{t}\hat{\phi}} |\tau_{12}|^{-p+\gamma_{\hat{\phi}}} \hat{\phi}(\tau_2) +O(|\tau_{12}|^0)\,.
\eal
Here, $\mathbb{1}$ denotes the identity operator, which is removed by subtracting the disconnected correlators. The $\hat{\phi}$ term introduces divergences upon integration over $\tau_1$, requiring a more careful treatment of it as a distribution. Explicitly, for an arbitrary smooth test function $\varphi(\tau)$, we have
\bal\label{general pole removal}
\int\limits_{|\tau|\leq 1} d^p \tau |\tau|^{-p+\gamma_{\hat{\phi}}} \varphi(|\tau|) =\vol_{S^{p-1}} \left(\frac{\varphi(0)}{\gamma_{\hat{\phi}}} +\int_0^1 d\tau |\tau|^{-1+\gamma_{\hat{\phi}}} (\varphi(\tau) -\varphi(0)) \right)\,.
\eal
Here we take $\varphi(\tau)=|\tau|^{-p-\gamma_{\hat{\phi}}} F_{2c}\left(\frac{\tau^2}{1+\tau^2}\right)$, in the limit $\tau\rightarrow 0$ it reduces to
\bal
\label{OPElimit}
\lim\limits_{\tau\rightarrow 0} |\tau|^{-p-\gamma_{\hat{\phi}}} F_{2c}\left(\frac{\tau^2}{1+\tau^2}\right) =b_{\cO_J \hat{\phi}} \lambda_{\hat{t}\hat{t}\hat{\phi}}\,.
\eal
This limit can be seen from inserting the $\hat{t}\times\hat{t}$ OPE into the connected three-point function \eqref{uOtt}---the contribution from the identity operator is canceled by the term $\langle\cO\rangle\langle\hat{t}\hat{t}\rangle$ in \eqref{disconnectedOtt}, while it is straightforward to see that any higher order terms in the OPE vanish as $|\tau|\rightarrow0$.
Subtracting this $\varphi(0)$ contribution renders the remaining integral on the right-hand side of \eqref{intF2} finite and straightforward to evaluate. Moreover, the integrand should be expanded in powers of $\gamma_{\hat{\phi}}$ perturbatively, leaving the integration measure as a sum of terms of the form $(\log^n \tau)/\tau$. Applying this procedure, \eqref{intF2} takes the modified form
\begin{multline}
\label{perturbativeF2int}
\frac{1}{2}\vol_{S^{p-1}} \left( \int_0^{1/2} d\chi \left(\frac{F_{2c}(\chi)}{\chi^{1+p/2} (1-\chi)^{1-p/2}} -\frac{b_{\cO_J \hat{\phi}} \lambda_{\hat{t}\hat{t}\hat{\phi}}}{\chi^{1-\gamma_{\hat{\phi}}/2} (1-\chi)^{1+\gamma_{\hat{\phi}}/2} }\right) +2\frac{b_{\cO_J \hat{\phi}} \lambda_{\hat{t}\hat{t}\hat{\phi}}}{\gamma_{\hat{\phi}}} \right)\\
+\frac{1}{2}\vol_{S^{p-1}} \int_{1/2}^1 d\chi \frac{F_{2c}(\chi)}{\chi^{1+p/2} (1-\chi)^{1-p/2}} =-J b_{\cO_J \hat{t}}\,.
\end{multline}
We again emphasize that the term $\frac{b_{\cO_J \hat{\phi}} \lambda_{\hat{t}\hat{t}\hat{\phi}}}{\chi^{1-\gamma_{\hat{\phi}}/2} (1-\chi)^{1+\gamma_{\hat{\phi}}/2} }$ should be expanded perturbatively to the order at which one has calculated $F_{2c}(\chi)$ in order to remove the pole at $\chi\rightarrow0$ and leave an integrable expression.

This structure becomes particularly elegant in certain cases---for instance, for the $1/2$ BPS Wilson loop in 
$\cN=4$ super Yang--Mills theory---where only the contact term contributes at leading order, yielding a remarkably simple expression
\bal\label{SYM Ott contact term leading order}
\vol_{S^{p-1}} \frac{b_{\cO_J \hat{\phi}} \lambda_{\hat{t}\hat{t}\hat{\phi}}}{\gamma_{\hat{\phi}}} \Bigg|_{\text{leading }} =-J b_{\cO_J \hat{t}}\Big|_{\text{leading }}\,.
\eal

Furthermore, the integrated $F_1$ term in \eqref{intF1} at leading order also encodes simple and universal relations among CFT data. At this order, $F_1(\chi)$ reduces to a constant,
\bal
\label{F1leading}
F_1(\chi)|_{\text{leading}}= \lambda_{\hat{t}\hat{t}\hat{T}} b_{\cO_J \hat{T}} \Big|_{\text{leading}}\,,
\eal
where $\hat{T}$ is the rank-2 traceless-symmetric operator appearing in the OPE of two tilts,
\bal
\hat{t}_i (\tau_1) \times \hat{t}_j (\tau_2) \ni \frac{1}{2} \left(\delta_{i}^k \delta_{j}^l +\delta_{i}^l \delta_{j}^k -\frac{2}{N-1}\delta_{ij} \delta^{kl} \right) \lambda_{\hat{t}\hat{t}\hat{T}} |\tau_{12}|^{\gamma_{\hat{T}}} \hat{T}_{kl}(\tau_2) \,.
\eal
Plugging \eqref{F1leading} into \eqref{intF1}, we obtain another very simple expression at leading order
\bal\label{topological sector leading order}
\vol_{S^{p-1}} \frac{\Gamma \left(\frac{p}{2}\right)^2}{2 \Gamma (p)} \lambda_{\hat{t}\hat{t}\hat{T}} b_{\cO_J \hat{T}} \Bigg|_{\text{leading}} &=J(J-1) b_{\cO_J \hat{t}} \Big|_{\text{leading}}\,.
\eal

Finally, let us say a few words about the case where $[\mathfrak{h}^{\perp},\mathfrak{h}^{\perp}]\nsubseteq \mathfrak{h}$. While $l_1$ itself is scheme-dependent, its commutator, as implied by \eqref{initiallambdaalgebra}, gives rise to a scheme-independent relation.
\begin{multline}
\rho^{\beta}(e_i,e_{\alpha}) \langle \cO_{\beta}(x_0) \hat{t}_j(\tau_2)\rangle_c -\rho^{\beta}(e_j,e_{\alpha}) \langle \cO_{\beta}( x_0) \hat{t}_i(\tau_2)\rangle_c \\
+ \left(\int d^p\tau_1 -\int d^p\tau_2 \right) \langle \cO_{\alpha}( x_0) \hat{t}_i(\tau_1) \hat{t}_j(\tau_2) \rangle_c + [e_i,e_j]^s \langle \cO_{\alpha} (x_0) \hat{t}_s(\tau_2) \rangle_c =0 \,.
\end{multline}
It is straightforward to check that the terms symmetric in the indices $i$ and $j$ cancel, so that only the antisymmetric component remains.

\subsubsection{\texorpdfstring{$\boldsymbol{n=2}$}{n=2}}
\label{sec:m=1n=2}
\begin{multline}
\langle \cO(\rho(\lambda,r),x_0) \hat{t}(w)^2\rangle_c + \langle \cO(r,x_0) \hat{t}(\lambda) \hat{t}^2 (w) \rangle_c + 2\langle \cO(r,x_0) \hat{t}(l_1(\lambda;w))\hat{t}(w) \rangle_c \\
 + \langle \cO(r) \hat{t}(l_2(\lambda;w,w))\rangle_c =0 \,.
\end{multline}
This is a slightly more involved case with three defect operators, where we again encounter the scheme-dependent term $l_2$, as in Section~\ref{sec:m=0n=3}. After taking the derivatives $\frac{\partial}{\partial \lambda^i} \frac{\delta}{\delta w^j (\tau_2)} \frac{\delta}{\delta w^k (\tau_3)}  \frac{\delta}{\delta r^{\alpha}(x_0)}$, we find
\bal
\label{m=1n=2der}
(T_i)_{\alpha}^{\beta} \langle \cO_{\beta}(x_0) \hat{t}_j (\tau_2) \hat{t}_k(\tau_3)\rangle_c + \int d^p \tau_1\langle \cO_{\alpha}(x_0) \hat{t}_i(\tau_1) \hat{t}_j (\tau_2) \hat{t}_k (\tau_3) \rangle_c&\\
+ l_1^s(e_i;e_j) \langle \cO_{\alpha}(x_0) \hat{t}_s(\tau_2) \hat{t}_k (\tau_3) \rangle_c
+ l_1^s (e_i;e_k) \langle \cO_{\alpha}(x_0) \hat{t}_s(\tau_3) \hat{t}_j (\tau_2) \rangle_c &\\
+ \delta^p (\tau_2-\tau_3) l_2^s(e_i;e_j,e_k) \langle \cO_{\alpha}(x_0) \hat{t}_s(\tau_3)\rangle_c =0\,.
\eal
Restricting to the case $[\mathfrak{h}^{\perp},\mathfrak{h}^{\perp}] \subseteq \mathfrak{h}$ where $l_1(e_i,e_j)=0$, similar to \eqref{nolog}, we can obtain a simple identity by choosing distinct insertion points. For instance, by setting $\tau_3 = 0$ and $\tau_2 = 1$ the result simplifies to 
\bal\label{m=1n=2SeparatePoints}
(T_i)_{\alpha}^{\beta} \langle \cO_{\beta}(x_0) \hat{t}_j (\tau_2) \hat{t}_k(\tau_3)\rangle_c + \int d^p \tau_1\langle \cO_{\alpha}(x_0) \hat{t}_i(\tau_1) \hat{t}_j (\tau_2) \hat{t}_k (\tau_3) \rangle_c =0\,.
\eal
This can be viewed as an analogue of \eqref{nolog} in the presence of a bulk operator. In order to extract the part associated with $l_2$ from \eqref{m=1n=2der}, we perform an additional integration over $\tau_2$ with a measure $\varphi(\tau_2)$, chosen to be the same as in \eqref{doubleintegraldefect}, i.e., $\varphi(0)=1$ and $\varphi(\infty)=0$. Upon this integration, the above relation simplifies to
\begin{multline}
(T_i)_{\alpha}^{\beta} \int d^p \tau_2 \varphi(\tau_2) \langle \cO_{\beta}(x_0) \hat{t}_j (\tau_2) \hat{t}_k(\tau_3)\rangle_c + \int d^p \tau_1 d\tau_2 \varphi(\tau_2) \langle \cO_{\alpha}(x_0) \hat{t}_i(\tau_1) \hat{t}_j (\tau_2) \hat{t}_k (\tau_3) \rangle_c \\
+ \varphi(\tau_3) l_2^s(e_i;e_j,e_k) \langle \cO_{\alpha}(x_0) \hat{t}_s(\tau_3)\rangle_c =0 \,.
\end{multline}
The integral $\int d^p \tau_1 d\tau_2 \varphi(\tau_2) \langle \cO_{\alpha}(x_0) \hat{t}_i(\tau_1) \hat{t}_j (\tau_2) \hat{t}_k (\tau_3) \rangle_c$ involves two kinds of potentially singular regions. The first occurs when two tilt operators approach each other, which, as discussed in Section~\ref{sec:m=1n=1}, requires no special regularization. The second arises when all three tilt operators collide, with the corresponding OPE given by
\bal
\label{logdiv}
\hat{t}_i (\tau_1) \hat{t}_j (\tau_2) \hat{t}_k (\tau_3) \ni r^{-2p} \lambda^n_{ijk}(y_1,y_2) \hat{t}_n(\tau_3) +O(r^{-2p+1})\,,
\eal
where $y_a=(\tau_a-\tau_3)/r$ encodes the angular dependence around the coincident point, and $r$ is the radial coordinate. The coefficients $\lambda^n_{ijk}$ can be read off from the four-point functions of the tilt operators. This term may induce a logarithmic divergence, requiring regularization in this limit via an appropriate renormalization scheme, as in Section~\ref{sec:m=0n=3}. Given the presence of $l_2$, a natural way to construct scheme-independent identities is to mimic \eqref{logid}, in which the commutators of two $l_2$’s eliminate the scheme-dependent parts,
\bal
(T_i)_{\alpha}^{\beta} \int d^p \tau_2 \varphi(\tau_2) \langle \cO_{\beta}(x_0) \hat{t}_j (\tau_2) \hat{t}_k(0)\rangle_c -(T_j)_{\alpha}^{\beta} \int d^p \tau_2 \varphi(\tau_2) \langle \cO_{\beta}(x_0) \hat{t}_i (\tau_2) \hat{t}_k(0)\rangle_c & \\
+ \int d^p \tau_1 d\tau_2 (\varphi(\tau_2) -\varphi(\tau_1)) \langle \cO_{\alpha}(x_0) \hat{t}_i(\tau_1) \hat{t}_j (\tau_2) \hat{t}_k (0) \rangle_c
+ [[e_i,e_j],e_k]^s \langle \cO_{\alpha}(x_0) \hat{t}_s(0)\rangle_c & =0 \,,
\eal
where for simplicity we set $\tau_3=0$. In contrast with \eqref{logid}, here the residual conformal group is too small to fully gauge away one integration variable in the double integral. Consequently, the result retains an explicit dependence on the measure $\varphi(\tau)$ and cannot be reduced to a $\varphi$-independent form like in \eqref{logexpression}. Nonetheless, the above expression holds for any choice of $\varphi(\tau)$ and in any renormalization scheme.

The expression for $l_2$ derived in the pure defect setup, \eqref{l2}, suggests an alternative route to eliminating the scheme dependence. Upon substituting \eqref{l2}, the resulting integral naturally couples pure defect correlators to those involving both bulk and defect insertions,
\begin{multline}
\int d^p \tau_1 d^p \tau_2 \varphi(\tau_2) (\langle \cO_{\alpha} (x_0) \hat{t}_i(\tau_1) \hat{t}_j (\tau_2) \hat{t}_k (0) \rangle_c-\langle \hat{t}_i (\tau_1) \hat{t}_j (\tau_2) \hat{t}_k (0) \hat{t}^s (\infty) \rangle_c  \langle \cO_{\alpha}(x_0) \hat{t}_s(0)\rangle_c )  \\
+ (T_i)_{\alpha}^{\beta} \int d^p \tau_2 \varphi(\tau_2) \langle \cO_{\beta}(x_0) \hat{t}_j (\tau_2) \hat{t}_k(0)\rangle_c = 0 \,.
\end{multline}
One of the advantages of this expression is that the potential logarithmic divergences \eqref{logdiv} in the 1-bulk–3-tilt correlator and the tilt four-point function cancel each other exactly, so no regularization is required.

\subsection{\texorpdfstring{$\boldsymbol{m\ge2}$}{mge2}}
Similar constraints can, of course, be constructed for the bulk–defect correlators in
\eqref{generalintid} with additional bulk insertions. Although the complexity grows rapidly
with the number of bulk operators $m$, bulk two-point functions are known in several
examples \cite{Chen:2023yvw, Gimenez-Grau:2022ebb, Barrat:2020vch, Bianchi:2022sbz}. An
advantage of bulk two-point (and higher-point) functions is that they admit conformal block
expansions in both the bulk and defect channels, allowing one to study them using bootstrap
techniques \cite{Bianchi:2022sbz}. In this context, the identities in \eqref{generalintid}
become a powerful tool: they can be incorporated into bootstrap algorithms—both analytical
and numerical—to generate genuinely new constraints on the CFT data.

In the previous sections, we have already demonstrated how the $m=1$ identity constrains the
CFT data, see \eqref{tiltwardid}, \eqref{perturbativeF2int}, \eqref{SYM Ott contact term
leading order}, and \eqref{topological sector leading order}. We expect that the identities
with $m \ge 2$ will impose additional, independent constraints, and we leave a detailed
analysis of these higher-order relations for future work.

In particular, there are examples in which bulk–defect correlators are easier to compute than
their purely bulk counterparts. A notable case is the magnetic line in the $O(N)$ model, where
the mixed correlator $\langle \phi \phi \hat{t} \hat{t} \rangle$ can be obtained at order
$\varepsilon$ in a relatively straightforward manner, since the relevant Feynman diagram is
the scalar box integral \eqref{Bloch-Wigner}. In contrast, the bulk two-point function
$\langle \phi \phi \rangle$ at the same order is substantially more complicated: its expression involves the
infinite-series representation \cite{Gimenez-Grau:2022ebb}. Our integral identities
therefore provide an alternative route to determining $\langle \phi \phi \rangle$ starting
from the simpler mixed correlator $\langle \phi \phi \hat{t} \hat{t} \rangle$,
\bal
2 \langle \phi_{\imath} (x_1) \phi_{\jmath} (x_2) \rangle_c -2\delta_{\imath \jmath} \langle \phi_{N} (x_1) \phi_{N} (x_2) \rangle_c =\int d\tau_3 d\tau_4 \langle \phi_N(x_1) \phi_N(x_2) \hat{t}_{\imath} (\tau_3)  \hat{t}_{\jmath} (\tau_4) \rangle_c\,,
\eal
for the explanation of field contents, see Section~\ref{sec:O(N)}. Even though, in this
specific example, the double tilt integral is as complicated to the direct
diagrammatic evaluation of $\langle \phi \phi \rangle$, the identity nonetheless illustrates
the broader potential of our method: it provides a universal mechanism
for relating bulk and bulk–defect observables, which may offer computational advantages in
other settings.

\section{\texorpdfstring{Example: 1/2 BPS Wilson Lines in $\cN=4$ SYM}{Example: 1/2 BPS Wilson Lines in N=4 SYM}}
\label{sec:N=4SYM}
In this section, we illustrate the integral identities derived above using the example of the $1/2$ BPS Wilson loop in $\cN=4$ super Yang--Mills theory~\cite{Maldacena:1998im, Drukker:1999zq}, which takes the form
\bal
\cW =\frac{1}{N} \tr \cP \exp \int \left( i A_{\mu} \dot{x}^{\mu} +|\dot{x}| \theta\cdot \phi \right) d\tau\,,
\eal
where we set $\theta^i=\delta^{i6}$, The remaining five scalars then play the role of the tilt operators appearing in~\eqref{WardJandt}~\cite{Alday:2007he, Polchinski:2011im, Beccaria:2017rbe, Bruser:2018jnc, Cuomo:2021rkm,Giombi:2017cqn}, with normalization
\bal
    C_{\hat{t}} = \frac{\sqrt{\lambda}}{2\pi^2} \frac{ I_2(\sqrt{\lambda}) }{ I_1(\sqrt{\lambda}) }\,,
\eal
where $C_{\hat{t}}$ is defined in \eqref{Zmetric}, and $I_n$ are the modified Bessel functions and $\lambda = N g_{YM}^2$ is the 't Hooft coupling. The bulk operators we consider are the single-trace operators, constructed similarly to \eqref{cOJ}, but with a constant prefactor:
\bal
\label{Giombibulk}
\cO_J (u,x)= (2\pi)^J \frac{2^{J/2}}{\sqrt{J \lambda^{J}}} \tr \left( u\cdot \phi(x) \right)^J
\eal
Here again we consider the case $J\ge 2$. The null vector $u$ guarantees that $\cO_J $ transforms in a rank-$J$ symmetric--traceless representation of the R-symmetry group. These operators are protected and have scaling dimensions equal to their R-charge $J$.

\subsection{\texorpdfstring{$\int\langle \cO \hat{t} \rangle \sim \langle \cO \rangle$}{m=1, n=0}}

The first identity that we want to demonstrate in this theory is \eqref{tiltwardid1d}, which can be rewritten simply as
\bal\label{N=4 a~b all couplings}
\frac{1}{\pi} a_{\cO_J} = b_{\cO_J \hat{t}}\,,
\eal
for the bulk operator $\cO_J$ defined in \eqref{Giombibulk}\footnote{Note that the definition of $b_{\cO_J \hat{t}}$ in \eqref{b_(OJ t)} is smaller by a factor of $1/J$ compared to the coefficient denoted 
$b_{\cO_J \hat{t}}$ in \cite{Artico:2024wnt}.}. The relevant CFT data entering this relation---the bulk one-point function and the bulk--tilt two-point function—were computed in~\cite{Giombi:2018hsx, Artico:2024wnt}:
\begin{align}
    \label{a_OJ}a_{\cO_J} & = \frac{\sqrt{\lambda J}}{N 2^{\frac{J}{2}+1}}\frac{I_J (\sqrt{\lambda})}{I_1(\sqrt{\lambda})}\,,\\
    \label{b_(OJ t)}b_{\cO_J \hat{t}} & = \frac{\sqrt{\lambda J} }{N 2^{\frac{J}{2}+1}\pi} \frac{ I_J(\sqrt{\lambda}) }{ I_1(\sqrt{\lambda}) }\,,
\end{align}
with $N$ the rank of the $SU(N)$ gauge group in $\cN=4$ super Yang--Mills.
It is straightforward to verify that the relation holds exactly: with the additional factor of $J/\pi$, the two coefficients match identically at all orders in the ’t~Hooft coupling~$\lambda$.

For later use, we expand $b_{\cO_J \hat{t}}$ and $a_{\cO_J} C_{\hat{t}}$ in the weak-coupling limits to the first few orders:
\begin{align}
\label{expanedb}
b_{\cO_J \hat{t}}&= \frac{ \sqrt{J}}{N 2^{3J/2}\pi \Gamma(J+1)} \lambda^{J/2} -\frac{\sqrt{J} (J-1)}{N 2^{3J/2+3} \pi \Gamma(J+2)} \lambda^{J/2+1}  +O(\lambda^{J/2+2})\,, \\
\label{expanedaC}
a_{\cO_J} C_{\hat{t}}&= \frac{\sqrt{J} }{N 2^{3J/2 +3} \pi ^2 \Gamma (J+1)} \lambda ^{J/2+1} -\frac{ (2 J-1) \sqrt{J} }{N 2^{3J/2+5} 3 \pi ^2 \Gamma (J+2)} \lambda ^{J/2+2} +O(\lambda ^{J/2+3})\,.
\end{align}

\subsection{\texorpdfstring{$\int\langle \cO \hat{t} \hat{t} \rangle \sim \langle \cO \hat{t} \rangle$}{m=1, n=1}}
In this section, our analysis is restricted to the leading and next-to-leading orders at both weak and strong coupling. 

In the $p=1$ case, equation \eqref{uOtt} becomes\footnote{Note that our $F_2$ is defined in a different way from that of \cite{Artico:2024wnt} by an overall factor of $-1/2$.}
\bal
\label{uOttp=1}
{}&\langle \cO_J(u, x_0) (\hat{u}_1 \cdot t)(\tau_1) (\hat{u}_2 \cdot t)(\tau_2) \rangle \\
&= \frac{(u\cdot \theta)^{J-2}}{|x_{0\perp}|^{\Delta_{\cO_J}-2}} \left( \frac{(u\cdot \hat{u}_1) (u\cdot \hat{u}_2)}{|x_{01}|^{2} |x_{02}|^{2}} F_{1}(\chi) +\frac{(u\cdot \theta)^2 (\hat{u}_1 \cdot \hat{u}_2)}{|x_{0\perp}|^2 |\tau_{12}|^2} F_{2}(\chi) \right)\,,
\eal
where the cross-ratio is $\chi = \frac{|x_{0\perp}|^2 |\tau_{12}|^2}{|x_{01}|^2 |x_{02}|^2}$ in \eqref{chi cross ratio}. Note that this is the full correlator rather than the connected part. Their relation is given by $F_{1c}=F_1$ and $F_{2c}=F_2 - a_{\cO_J} C_{\hat{t}}$, as discussed in Section~\ref{sec:m=1n=1}.

In particular, the pinching and splitting limits introduced in \cite{Artico:2024wnt} provide non-perturbative constraints on these functions. The pinching limit corresponds to taking $\tau_1\rightarrow \tau_2$, where the defect insertions approach each other and the correlator reduces to a lower-point function along the Wilson loop. A different limit comes from considering a large separation between the bulk operator and the defect operators. In the correlator expressed in R-symmetry channels, this corresponds to the limit $|x_{0\perp}|\rightarrow \infty$. In this case, the bulk–defect–defect correlator factorizes into a product of a two-point defect correlator and a one-point function of the bulk operator. Note that both of these limits cause $\chi\rightarrow0$---applying them leads to the exact relations
\bal
\label{pinchingsplittinglimits}
F_1(0)=b_{\cO_J \hat{T}}\lambda_{\hat{t}\hat{t}\hat{T}} ,\qquad\qquad F_2(0)=a_{\cO_J} C_{\hat{t}}\,,
\eal
where at weak coupling the first few orders read
\bal
\label{bOJTlambdattT}
b_{\cO_J \hat{T}}\lambda_{\hat{t}\hat{t}\hat{T}}=\frac{\sqrt{J} \lambda ^{J/2}}{N 2^{3 J/2} \pi ^2 \Gamma (J-1)} -\frac{ \lambda ^{J/2+1}}{N 2^{3J/2+3 } \pi ^2 \sqrt{J} \Gamma (J-2)} +O(\lambda ^{J/2+2})\,.
\eal
We note that the coefficients $b_{\cO_J \hat{T}}$ and $\lambda_{\hat{t}\hat{t}\hat{T}}$ are given
in closed form to all orders in \cite{Artico:2024wnt}.

A further relation between $F_{1}$ and $F_{2}$ comes from the topological sector. In $\cN=4$ super Yang--Mills, the 1/2 BPS Wilson line is known to possess a topological sector \cite{Giombi:2018hsx, Giombi:2018qox, Beem:2013sza,Drukker:2007yx, Beccaria:2020ykg}, consisting of correlators of 1/2 BPS operators that become effectively topological when the spacetime and R-symmetry polarizations are aligned. In this limit, the correlators lose their dependence on continuous spacetime separations and depend only on discrete data such as operator ordering or internal symmetry structure. In our setup, we focus on the bulk–tilt–tilt correlator \cite{Artico:2024wnt}
\bal
\label{topF12}
\bF_J =F_1(\chi)-2 F_2(\chi)\,,
\eal
which reflects the decomposition of topological sector into R-symmetry channels. 

This topological sector admits an equivalent interpretation in terms of the OPE of defect operators. A easy way to see it is
\bal
\bF_J =F_1(0)-2 F_2(0)=b_{\cO_J \hat{T}}\lambda_{\hat{t}\hat{t}\hat{T}} -2 a_{\cO_J} C_{\hat{t}}\,.
\eal
Combining this representation with the general form above, one can write 
\bal
\label{F1relation}
F_1(\chi)= b_{\cO_J \hat{T}}\lambda_{\hat{t}\hat{t}\hat{T}} -2 a_{\cO_J} C_{\hat{t}} +2 F_2(\chi)\,.
\eal

\subsubsection{Weak coupling}
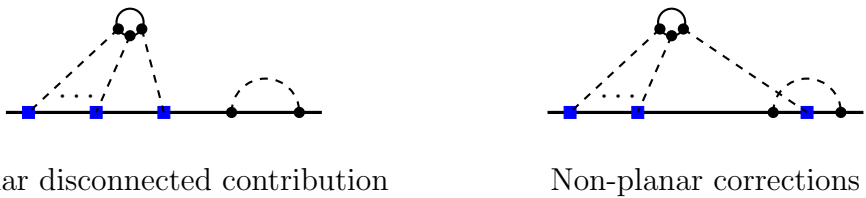
\begin{figure}[ht!]
\centering
\begin{tikzpicture}[scale=0.6]

\tikzstyle{blackdot}=[circle, fill=black, inner sep=1.5pt]
\tikzstyle{bluesquare}=[draw=blue, fill=blue, minimum size=4.5pt, inner sep=0pt]
\tikzstyle{bottomline} = [very thick]

\coordinate (L1) at (0,0);
\coordinate (L2) at (1.5,0);
\coordinate (L3) at (3,0);
\coordinate (L4) at (4.5,0);
\coordinate (L5) at (6,0);

\path (L1) -- (L2) coordinate[midway] (Lellip);

\draw[bottomline] (-0.5,0) -- (6.5,0);

\node[above=2mm, right=-2mm] at (Lellip) {$\cdots$};

\coordinate (C1) at (2.25,2);
\def\r{0.3}
\draw[thick] (C1) circle (\r);

\foreach \angle/\label in {210/a, 270/b, 330/c} {
    \path (C1) ++(\angle:\r) coordinate (P\label);
    \node[blackdot] at (P\label) {};
}

\node[bluesquare] at (L1) {};
\node[bluesquare] at (L2) {};
\node[bluesquare] at (L3) {};

\draw[thick, dashed] (Pa) -- (L1);
\draw[thick, dashed] (Pb) -- (L2);
\draw[thick, dashed] (Pc) -- (L3);

\draw[thick, dashed] (L4) arc (180:0:0.75);

\node[blackdot] at (L4) {};
\node[blackdot] at (L5) {};

\node[below=6mm] at (3,0) {\text{Planar disconnected contribution}};

\begin{scope}[xshift=12cm]  

\coordinate (R1) at (0,0);
\coordinate (R2) at (1.5,0);
\coordinate (R3) at (3,0);
\coordinate (R4) at (4.5,0);
\coordinate (R5) at (6,0);

\path (R1) -- (R2) coordinate[midway] (Rellip);
\path (R4) -- (R5) coordinate[midway] (Rmid);

\draw[bottomline] (-0.5,0) -- (6.5,0);

\node[above=2mm, right=-2mm] at (Rellip) { $\cdots$};

\coordinate (C2) at (2.25,2);
\draw[thick] (C2) circle (\r);

\foreach \angle/\label in {210/a, 270/b, 330/c} {
    \path (C2) ++(\angle:\r) coordinate (Q\label);
    \node[blackdot] at (Q\label) {};
}

\node[bluesquare] at (R1) {};
\node[bluesquare] at (R2) {};
\node[bluesquare] at (Rmid) {};

\draw[thick, dashed] (Qa) -- (R1);
\draw[thick, dashed] (Qb) -- (R2);
\draw[thick, dashed] (Qc) -- (Rmid);

\draw[thick, dashed] (R4) arc (180:0:0.75);

\node[blackdot] at (R4) {};
\node[blackdot] at (R5) {};

\node[below=6mm] at (3,0) {\text{Non-planar corrections}};

\end{scope}

\end{tikzpicture}%
\caption{Disconnected Feynman diagrams contributing to $F_2$ in $\langle\cO_J tt\rangle$ correlators at weak coupling. The ellipsis indicates the $J-3$ propagators connecting the bulk operator to the Wilson line that have been omitted for clarity. The black circles represent bulk field insertions, while the blue squares denote the scalar coupling $\phi^6$ that are
integrated along the Wilson loop.}
\label{fig:DisconnectedFeynmandiagrams}
\end{figure}

We now focus on the weak-coupling regime. At leading order in $\lambda$, the function $F_2$ does not contribute. At next-to-leading order, it receives contributions only from disconnected Feynman diagrams, as illustrated in Figure~\ref{fig:DisconnectedFeynmandiagrams}. The left diagram represents planar disconnected contributions that survive in the large-$N$ limit, while the right diagram corresponds to non-planar contributions that vanish in this limit. Although these diagrams are disconnected, $F_2$ is not strictly constant: the integration over the positions of the Wilson line insertions must be treated carefully, as the integration region and ordering of the propagators can introduce a nontrivial dependence on the cross ratio $\chi$. 

The planar contribution from the left diagram has been calculated in \cite{Artico:2024wnt} and is given by
\bal
\label{F2}
F_2(\chi)=\lambda^{J/2+1} c_{J} \sum_{\pm}\left(\pi\pm 2\arctan \sqrt{\frac{1-\chi}{\chi} } \right)^J +O(\lambda^{J/2+2})\,,
\eal
where $c_J$ is a constant that is fixed using the pinching and splitting limits \eqref{pinchingsplittinglimits}. Inserting the data from \eqref{expanedaC}, we obtain
\bal
c_{J}= \frac{\sqrt{J} }{N 2^{\frac{5}{2} J+3}\pi ^{J+2} \Gamma (J+1)}\,.
\eal

\begin{figure}[ht!]
    \centering
    \begin{tikzpicture}[scale=0.6]

\tikzstyle{blackdot}=[circle, fill=black, inner sep=1.5pt]
\tikzstyle{bluesquare}=[draw=blue, fill=blue, minimum size=4.5pt, inner sep=0pt]
\tikzstyle{bottomline} = [very thick]

\coordinate (L0) at (-1.5,0); 
\coordinate (L1) at (0,0);    
\coordinate (L2) at (1.5,0);  
\coordinate (L3) at (3,0);    

\draw[bottomline] (-1.8,0) -- (3.5,0);

\coordinate (C1) at (2.25,2);
\def\r{0.3}
\draw[thick] (C1) circle (\r);

\foreach \angle/\label in {210/a, 270/b, 330/c} {
    \path (C1) ++(\angle:\r) coordinate (P\label);
    \node[blackdot] at (P\label) {};
}

\node[bluesquare] at (L0) {}; 
\node[bluesquare] at (L1) {}; 
\node[blackdot] at (L2) {};
\node[blackdot] at (L3) {};

\draw[thick, dashed] (Pa) -- (L0); 
\draw[thick, dashed] (Pa) -- (L1); 
\draw[thick, dashed] (Pb) -- (L2);
\draw[thick, dashed] (Pc) -- (L3);

\path (L0) -- (L1) coordinate[midway] (Lellip);
\node[above=2mm,right=0.2mm] at (Lellip) {$\cdots$};

\end{tikzpicture}
    \caption{Feynman diagram contributing at leading order in the $F_1$ channel.}
    \label{fig:FDF1}
\end{figure}
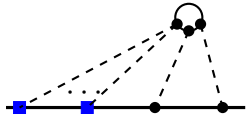

Since $F_{2}$ vanishes at the leading order $\lambda^{J/2}$, we turn to the $F_{1}$ channel,
which already contributes at order $\lambda^{J/2}$ as a constant. This follows from the
Feynman diagram analysis shown in Figure~\ref{fig:FDF1}, and is consistent with
\eqref{F1relation}.
From \eqref{expanedaC}, we see that the term proportional to $a_{\mathcal{O}_J}C_{\hat{t}}$ in \eqref{F1relation} is also sub-leading. Therefore, we have
\bal
F_1(\chi)= \lambda^{J/2} (b_{\cO_J \hat{T}}\lambda_{\hat{t}\hat{t}\hat{T}})|_{\lambda^{J/2}} +O(\lambda^{J/2+1})\,.
\eal

From \eqref{topological sector leading order} for $p=1$, we have
\bal
\pi \lambda_{\hat{t}\hat{t}\hat{T}} b_{\cO_J \hat{T}}\Big|_{\text{leading}} &=J(J-1) b_{\cO_J t} \Big|_{\text{leading}}\,.
\eal
which leads to the explicit solution
\bal
\label{leadingbOT}
b_{\cO_J \hat{T}}\lambda_{\hat{t}\hat{t}\hat{T}} =\frac{\sqrt{J}}{N 2^{3J/2}\pi^2 \Gamma(J-1)} \lambda^{J/2} +O(\lambda^{J/2+1})\,,
\eal
in agreement with \eqref{bOJTlambdattT} at leading order. This provides a consistency check,
confirming that our results align with previously computed expressions.

At the next-to-leading order, the Feynman diagrams contributing to the $F_1$ channel become significantly more involved. However, we can bypass the explicit diagrammatic computation by making use of \eqref{F1relation}, which provides a shortcut to determine $F_1$ at this order without evaluating the full set of diagrams. Specifically, substituting \eqref{F1relation} into \eqref{intF1} for $p=1$ yields an integral equation for $F_{2}$:
\bal
\pi (b_{\cO_J \hat{T}}\lambda_{\hat{t}\hat{t}\hat{T}} -2 a_{\cO_J} C_{\hat{t}}) + 2 \int_0^1 \frac{d\chi\, F_2(\chi)}{(\chi(1-\chi))^{1/2}} &=J(J-1) b_{\cO_J t}\,.
\eal
Using \eqref{F2}, \eqref{bOJTlambdattT}, \eqref{expanedb}, and \eqref{expanedaC}, we verified
that this relation is satisfied exactly at order $\lambda^{J/2+1}$. This provides a
nontrivial check of the expression for $F_{2}$ given in \eqref{F2}.

Now we turn to the $F_2$ channel and we emphasize that the disconnected diagrams are not equivalent to the disconnected correlators. As discussed in \eqref{disconnectedOtt} and in Appendix~\ref{sec:connected}, the disconnected correlators are defined by subtracting the factorized contribution from the full correlator. In the present case, this corresponds to removing the product of the bulk one-point function and the tilt two-point function. Explicitly, this yields \eqref{disconnectedF}. Plugging it into the identity \eqref{perturbativeF2int}, we find
\begin{multline}\label{F2 N=4 relation}
2\frac{b_{\cO_J \hat{\phi}} \lambda_{\hat{t}\hat{t}\hat{\phi}}}{\gamma_{\hat{\phi}}} +\int_0^{1/2} d\chi \left(\frac{F_{2c}(\chi)}{\chi^{3/2} (1-\chi)^{1/2}} -\frac{b_{\cO_J \hat{\phi}} \lambda_{tt\hat{\phi}}}{\chi^{1-\gamma_{\hat{\phi}}/2} (1-\chi)^{1+\gamma_{\hat{\phi}}/2} }\right)+\int_{1/2}^1 d\chi \frac{F_{2c}(\chi)}{\chi^{3/2} (1-\chi)^{1/2}}\\
=-\frac{\sqrt{J}}{N 2^{3J/2}\pi \Gamma(J)} \lambda^{J/2} +\frac{J^{3/2} (J-1)}{N 2^{3J/2+3}\pi \Gamma(J+2)} \lambda^{J/2+1}\,.
\end{multline}
Note again that the $\frac{1}{\chi^{1-\gamma_{\hat{\phi}}/2} (1-\chi)^{1+\gamma_{\hat{\phi}}/2} }$ term must be expanded perturbatively. Here, the left-hand side receives contributions from both the contact term and the integrated part of the disconnected correlator. However, at leading order in $\lambda$, the integral is subleading, and the only contribution comes from the contact term. Using \cite{Cavaglia:2022qpg}
\bal
\label{lambdattphi}
\lambda_{{\hat{t}\hat{t}\hat{\phi}}}=\frac{\lambda ^{3/2}}{16 \sqrt{2} \pi ^3}-\frac{\left(18+\pi ^2\right) \lambda ^{5/2}}{768 \sqrt{2} \pi ^5},\qquad \gamma_{\hat{\phi}}=\frac{\lambda }{4 \pi ^2} -\frac{\lambda ^2}{16 \pi ^4}+  \frac{1}{32\pi^2}\left( \frac{1}{\pi^4} - \frac{7}{720} \right) \lambda ^3 +O(\lambda^4)\,,
\eal
we find
\bal
\label{expectedbOphi}
b_{\cO_J \hat{\phi}} =-\frac{\sqrt{J}}{N 2^{3(J-1)/2}\Gamma(J)} \lambda^{(J-1)/2} +O(\lambda^{(J+1)/2})\,.
\eal
The same result may be extracted directly from the OPE limit \eqref{OPElimit}. 
Indeed,
\bal
(b_{\cO_J \hat{\phi}} \lambda_{\hat{t}\hat{t}\hat{\phi}})|_{\lambda^{J/2+1}} =\lim\limits_{\tau\rightarrow 0} \frac{1}{\tau} F_{2c} \left(\frac{\tau^2}{1+\tau^2} \right) \bigg|_{\lambda^{J/2+1}}=-\frac{\sqrt{J}}{N 2^{3(J+2)/2} \pi ^3 \Gamma (J)}\,,
\eal
which matches the product of \eqref{lambdattphi} and \eqref{expectedbOphi}, providing a nontrivial consistency check. This generalizes (4.17) of \cite{Artico:2024wnt}, which corresponds to $J=2$.

At next-to-leading order, the integrals in \eqref{F2 N=4 relation} start to contribute.
After performing the integrals and making use of the result for $b_{\mathcal{O}_J \hat{\phi}}$ at leading order, which is given in \eqref{expectedbOphi}, we find that at next-to-leading order,
\begin{equation}
\label{bOJphi}
    b_{\mathcal{O}_J \hat{\phi}}\bigg|_{\lambda^{(J+1)/2}} \!\!\!\!\!= \frac{\sqrt{J}}{N 2^{3(J+1)/2}\Gamma(J)} \left( \frac{J-1}{J+1} + 2 \frac{1-\log2}{\pi^2} + \sum_{j=1}^{(J-2)/2} \frac{(-1)^j \Gamma(J-1+2j) \zeta(2j+1)}{2^{2j-1}\pi^{2j+2} \Gamma(J-1)} \right).
\end{equation}
To the best of our knowledge, this quantity has not been computed previously in the
literature, neither via localization nor via explicit Feynman diagram calculations. We
therefore present it as a prediction and as an illustration of how the integral identities
constrain the CFT data.

\subsubsection{Strong coupling}
At strong coupling, there is, as of now, relatively little that can be said in general terms. The leading and next-to-leading correlators were computed in \cite{Artico:2024wnt}. However, if we assume that $F_1(\chi)$ remains constant at both leading and next-to-leading orders as claimed in that paper, then using \eqref{intF1} we obtain
\bal
F_{1}(\chi)= \frac{(J-1) J^{5/2}}{\pi^2 2^{J/2+1}}\sqrt{\lambda} -\frac{J^{5/2}(J-1)^2 (J+1) }{\pi ^2 2^{J/2+2}} +O(\lambda^{-1/2})\,.
\eal
For $J=2$, this simplifies to
\bal
F_{1} (\chi)= \frac{\sqrt{2}}{\pi ^2} \sqrt{\lambda } -\frac{3}{\sqrt{2} \pi ^2} +O(\lambda^{-1/2})\,.
\eal
Comparing this with the results of \cite{Artico:2024wnt}---where we have accounted for the canonical normalization of the tilt operators, as well as the additional factor of $-2$ arising from our different definition of $F_2$\footnote{We thank Daniele Artico and Julien Barrat for updating $F_1$ and $F_2$ in \cite{Artico:2024wnt} at next-to-leading order.}---we find
\bal
F_{1}=-\frac{3}{\sqrt{2} \pi^2} +O(\lambda^{-1/2})\,,
\eal
which agrees with our expression at next-to-leading order but differs at leading order.

Moreover, at both leading and next-to-leading orders, $F_2(\chi)$ cannot be constant, contrary to the claim in \cite{Artico:2024wnt}. If $F_2(\chi)$ were constant, it would violate \eqref{intF2}, since in this regime there are no contact terms analogous to those appearing at weak coupling, and the integral relation \eqref{intF2} should therefore hold exactly without any subtleties or modifications. Furthermore, this observation implies that $F_1(\chi)$ cannot remain constant either, as the two channels are linked through the topological sector, see \eqref{topF12}. Taken together, these results indicate that, at both weak and strong coupling, the bulk--defect--defect correlators calculated in \cite{Artico:2024wnt} are inconsistent with the integral identities. We anticipate that our identities will provide a robust framework for checking and revising existing correlators, leaving a complete reassessment for future work.

\section{\texorpdfstring{Example: Magnetic Lines in $O(N)$}{Example: Magnetic Lines in O(N)}}
\label{sec:O(N)}

The bulk $O(N)$ model in Euclidean $d=4-\varepsilon$ dimensions consists of $N$ real scalar fields $\phi_{0a}$, $a=1,\dots,N$, governed by the action
\begin{equation}
    S_{O(N)} = \int d^d x \left( \frac{1}{2} \partial_\mu \phi_{0a} \partial^\mu \phi_{0a} + \frac{1}{4!} \lambda_0 \left( \phi_{0a} \phi_{0a} \right)^2 \right)\,,
\end{equation}
where the subscript $0$ denotes bare quantities.
The magnetic line, taken to lie along $x^\mu(\tau) = (\tau,0_\perp)$, is given by deforming the bulk action,
\begin{equation}
    S_{O(N)} \rightarrow S_{O(N)} - h_0 \int_{-\infty}^{\infty} d\tau \phi_{0N} (\tau)\,,
\end{equation}
where we use the $O(N)$ symmetry to take $\phi_{0N}$ as the direction which couples to the line---in the language of the polarization vectors, we take $\theta^a=\delta^{aN}$. This breaks the bulk $O(N)$ symmetry to $O(N-1)$---in the following, we label the unbroken directions by $\imath,\jmath=1,\dots,N-1$. When working perturbatively in $\varepsilon$, this theory exhibits an RG flow to an infrared fixed point which, to linear order, is characterized by \cite{Cuomo:2021kfm}
\begin{equation}
    \frac{\lambda_*}{(4\pi)^2} = \frac{3\varepsilon}{N+8} + O(\varepsilon^2) \,,\quad h_*^2 = N+8 + \frac{4N^2+45N+170}{2(N+8)}\varepsilon + O(\varepsilon^2)\,.
\end{equation}
The tilt operator $\hat{t}$ has protected scaling dimension $\Delta_{\hat{t}}=1$. The canonical normalization of the tilt is given by
\begin{equation}
    C_{\hat{t}} = \frac{h_*^2}{4\pi^2} \left(1-\frac{\varepsilon}{2} +O(\varepsilon^2)\right)\,.
\end{equation}

In the following sections we demonstrate a variety of integral identities each with a single bulk operator insertion, a tilt operator inserted on the defect, and either zero, one or two other defect operators. We primarily consider bulk operators with dimensions close to one and two and defect operators with dimensions close to one. All operators are unit-normalized with the exception of the tilt operator.

\subsection{\texorpdfstring{$\int\langle \cO \hat{t} \rangle \sim \langle \cO \rangle$}{m=1, n=0}}
In this section, we make use of the identities \eqref{m=1n=0General} and \eqref{tiltwardid1d} in order to determine the one-point function coefficient of a selection of low dimensional operators.
First, we consider $\cO=\phi_a$. Then $b_{\phi \hat{t}}$ has been calculated in \cite{Gimenez-Grau:2022ebb} to be 
\begin{equation}
    b_{\phi \hat{t}} = \sqrt{C_{\hat{t}}} \left( 1-\frac{1-\log 2}{2}\varepsilon +O(\varepsilon^2)\right) \,.
\end{equation}
Using this in \eqref{tiltwardid1d}, we find
\begin{equation}
    a_\phi = \frac{\sqrt{N+8}}{2} \left(1+\left(\log\sqrt{2} + \frac{N^2-3N-22}{4(N+8)^2} \right) \varepsilon\right)\,,
\end{equation}
which agrees with what is calculated in \cite{Gimenez-Grau:2022ebb} via Feynman diagrams.

We now take $\cO = S \equiv Z_S^{-1} \phi_a\phi_a $. Then since $\langle S(x_1)\hat{t}_\imath (\tau_2)\rangle = 0$ by symmetry considerations and $T_\imath S = 0$, the identity \eqref{m=1n=0General} is trivial and we do not find any constraint on the bulk one-point function coefficient $a_S$. This quantity is instead determined by Feynman diagrams and has been calculated in \cite{Gimenez-Grau:2022ebb}.

Finally, we take $\cO = T_{ab} \equiv Z_T^{-1} (\phi_a \phi_b -\tr)$. From a Feynman diagram calculation, we find
\begin{equation}
    \label{b_Tt} b_{T\hat{t}} = \frac{N+8}{4\sqrt{2}\pi} \left( 1+\left( \frac{N+6}{N+8}\log2 +\frac{N^2-5N-38}{2(N+8)^2} \right)\varepsilon \right)\,.
\end{equation}
In this case, \eqref{tiltwardid1d} gives
\begin{equation}
    a_T = \frac{N+8}{4\sqrt{2}}\left( 1+ \left( \frac{N+6}{N+8}\log2 + \frac{N^2-5N-38}{2(N+8)^2}\right)\varepsilon\right)\,,
\end{equation}
which also agrees with results in \cite{Gimenez-Grau:2022ebb}.

\subsection{\texorpdfstring{$\int\langle \cO \hat{\cO} \hat{t} \rangle \sim \langle \cO \hat{\cO} \rangle$}{m=1, n=0 with extra defect operator}}
We now consider the identities in \eqref{intF1} and \eqref{intF2} and manipulations thereof. We consider bulk operators with dimensions close to one or two. Moreover, in Section \ref{sec:derivation} we only considered tilt operators inserted on the defect. It is straightforward to generalize to other defect operators and here we consider situations where the dimension one singlet $\hat{\phi}$ is inserted.

\subsubsection{\texorpdfstring{Bulk Operator Dimension $\sim1$}{Bulk Operator Dimension sim 1}}

Let us now consider the integral identity that involves integrating $\langle \phi_a \hat{t}_\imath \hat{t}_\jmath \rangle_c$. We use conformal symmetry to place the operators at $x_0 = (0,1)$, $\tau_1 = 0$ and $\tau_2 \equiv \tau$ and integrate over $\tau$. From a Feynman diagram calculation, see \eqref{F_phi t t full}, we find that
\begin{equation}
    \label{phi t t connected} \langle (u\cdot\phi) (0,1) (\hat{u}_1\cdot\hat{t})(0) (\hat{u}_2\cdot\hat{t})(\tau) \rangle_c = (u\cdot\theta)(\hat{u}_1\cdot\hat{u}_2) \varepsilon\frac{\sqrt{N+8}}{16\pi^2} \frac{4\arccos^2 \sqrt{\chi}-\pi^2}{(1+\tau^2)\chi}\,,
\end{equation}
where the cross-ratio is given by $\chi=\frac{\tau^2}{1+\tau^2}$.
In the language of \eqref{uOtt}, we have 
\begin{equation}
    F_{2c}(\chi)=\varepsilon\frac{\sqrt{N+8}}{16\pi^2}(4\arccos^2\sqrt{\chi}-\pi^2)\,.
\end{equation}
We take $\varphi(\tau)$ as prescribed in the discussion below \eqref{general pole removal} to be $\varphi(\tau) = |\tau|^{-1-\gamma_{\hat{\phi}}}F_{2c}(\chi)$ which reduces to $\varphi(0) = b_{\phi \hat{\phi}} \lambda_{\hat{t}\hat{t}\hat{\phi}}$, where \cite{Sakkas:2024dvm,Cuomo:2021kfm}
\begin{align}
    \label{b phi phi} b_{\phi \hat{\phi}} & = 1-\frac{3-3\log2}{2}\varepsilon+O(\varepsilon^2)\,,\\
    \label{lambda t t phi} \lambda_{\hat{t}\hat{t}\hat{\phi}} & = -C_{\hat{t}} \left(\frac{\pi}{\sqrt{N+8}} \varepsilon -\frac{(9N^2 +127 N+494)\pi}{4(N+8)^{5/2}} \varepsilon^2 +O(\varepsilon^3)\right)\,,\\
    \label{Delta phi}\Delta_{\hat{\phi}} & = 1+\varepsilon-\frac{3N^2+49N+194}{2(N+8)^2} \varepsilon^2 +O(\varepsilon^3)\,.
\end{align}
Plugging these ingredients into \eqref{perturbativeF2int}, we find the left-hand side to be
\begin{multline}
    2\frac{b_{\phi\hat{\phi}} \lambda_{\hat{t}\hat{t}\hat{\phi}} }{\gamma_{\hat{\phi}}} + \int_0^\frac{1}{2}d\chi \left(\frac{F_{2c}(\chi)}{\chi^{3/2}(1-\chi)^{1/2}} - \frac{b_{\phi\hat{\phi}} \lambda_{\hat{t}\hat{t}\hat{\phi}}}{\chi (1-\chi) } \right) + \int_{\frac{1}{2}}^1 d\chi \frac{F_{2c}(\chi)}{\chi^{3/2}(1-\chi)^{1/2}}
    \\ = -\frac{\sqrt{N+8}}{2\pi} \left( 1 + \left( \frac{N^2-3N-22}{4(N+8)^2}+\frac{\log2}{2}\right)\varepsilon\right)\,.
\end{multline}
The identity \eqref{perturbativeF2int} allows us to determine $b_{\phi\hat{t}}$, which was previously calculated in \cite{Gimenez-Grau:2022ebb} using Feynman diagrams, and reads
\begin{equation}
    b_{\phi \hat{t}} = \sqrt{C_{\hat{t}}} \left( 1-\frac{1-\log2}{2}\varepsilon +O(\varepsilon^2)\right)\,.
\end{equation}

We now consider the case where we insert the defect operator $\hat{\phi}$ rather than a second tilt operator. We calculate the correlation function in \eqref{F_phi phi t full} from Feynman diagrams to find
\begin{align}
    \langle (u\cdot\phi)(0,1) \hat{\phi}(0) (\hat{u}_1 \cdot\hat{t})(\tau)\rangle_c & = (u\cdot \hat{u}_1)\frac{\varepsilon}{8\pi} \frac{4\arccos^2 \sqrt{\chi} - \pi^2 }{(1+\tau^2)\chi}\,.
\end{align}
Whereas before, the divergence of the integral from $\tau \rightarrow 0$ came from the $\hat{\phi}$ contribution in the $\hat{t} \times \hat{t}$ OPE, the divergence in the present case is from the $\hat{t}$ contribution to the $\hat{\phi}\times\hat{t}$ OPE. Taking $F_{2c}(\chi)=\frac{\varepsilon}{8\pi}(4\arccos^2\sqrt{\chi}-\pi^2)$, the identity analogous to \eqref{perturbativeF2int} that we find is
\begin{multline}
    -2\frac{C_{\hat{t}}^{-1} b_{\phi\hat{t}} \lambda_{\hat{t}\hat{t}\hat{\phi}} }{\gamma_{\hat{\phi}}} + \int_0^\frac{1}{2}d\chi \left(\frac{F_{2c}(\chi)}{\chi^{3/2}(1-\chi)^{1/2}} - \frac{C_{\hat{t}}^{-1} b_{\phi\hat{t}} \lambda_{\hat{t}\hat{t}\hat{\phi}}}{\chi(1-\chi) } \right) + \int_{\frac{1}{2}}^1 d\chi \frac{F_{2c}(\chi)}{\chi^{3/2}(1-\chi)^{1/2}}
    \\ = 1 - \frac{3-3\log2}{2}\varepsilon = b_{\phi \hat{\phi}}\,.
\end{multline}
We read off the expression for $b_{\phi \hat{\phi}}$ which is consistent with the calculation in \cite{Gimenez-Grau:2022ebb} using Feynman diagrams.

\subsubsection{\texorpdfstring{Bulk Operator Dimension $\sim2$}{Bulk Operator Dimension sim 2}}

Of the two bulk operators at this dimension, we first consider the singlet $S$. There is only one non-zero three point function involving $S$ and defect operators with dimensions close to one that can be integrated in our framework, $\langle S \hat{t} \hat{t} \rangle_c$. The full correlation function is given in \eqref{F_S t t full}. The connected three-point function is given by $\langle S \hat{t} \hat{t} \rangle_c = \langle S \hat{t} \hat{t} \rangle - \langle S \rangle \langle\hat{t} \hat{t}\rangle$, where the one-point function $\langle S \rangle$ has been calculated in \cite{Gimenez-Grau:2022ebb}. The result is $\langle S(0,1) (\hat{u}_1\cdot\hat{t}) (0) (\hat{u}_2\cdot\hat{t}) (\tau) \rangle_c = \frac{\hat{u}_1\cdot\hat{u}_2}{1+\tau^2} F_{2c}(\chi)$, where now
\begin{equation}
        F_{2c}(\chi) = \frac{(N+8)\chi}{2\pi^2\sqrt{2N}} \biggl( 1 + \varepsilon \biggl( -\frac{\pi^2-4\arccos^2 \sqrt{\chi} }{8\chi} -\frac{13N+38}{2(N+8)^2}+\log2+\frac{N+2}{2(N+8)}\log \chi\biggr) \biggr)\,.
\end{equation}
Based on symmetry considerations, $b_{S\hat{t}}=0$, which means that the integral identity takes the form
\begin{equation}
    \label{b_S phi constraint} 2 \frac{b_{S\hat{t}} \lambda_{\hat{t}\hat{t}\hat{\phi}}}{\gamma_{\hat{\phi}}} + \int_0^\frac{1}{2}d\chi \left(\frac{F_{2c}(\chi)}{\chi^{3/2}(1-\chi)^{1/2}} - \frac{b_{S\hat{\phi}} \lambda_{\hat{t}\hat{t}\hat{\phi}}}{\chi (1-\chi) } \right) + \int_{\frac{1}{2}}^1 d\chi \frac{F_{2c}(\chi)}{\chi^{3/2}(1-\chi)^{1/2}}=0\,.
\end{equation}
This equation can be used to determine $b_{S\hat{t}}$ as follows. The leading order part of $b_{S\hat{\phi}}$ can be calculated from considering \eqref{b_S phi constraint} at $O(\varepsilon^0)$---at this order, the term $\frac{b_{S\hat{\phi}} \lambda_{\hat{t}\hat{t}\hat{\phi}}}{\chi (1-\chi) }$ is zero and $\frac{F_{2c}(\chi)}{\chi^{3/2}(1-\chi)^{1/2}}$ is integrable over $\chi\in(0,1)$.
If we write $b_{S\hat{\phi}}=b_{S\hat{\phi}}^{(0)}+\varepsilon b_{S\hat{\phi}}^{(1)}$, then we find
\begin{equation}
    -b_{S\hat{\phi}}^{(0)} \frac{\sqrt{N+8}}{2\pi} + \frac{N+8}{2\pi\sqrt{2N}} +O(\varepsilon) = 0\,,
\end{equation}
from which we conclude $b_{S\hat{\phi}}^{(0)} = \sqrt{\frac{N+8}{2N}}$. Moving to the next-to-leading order, $\frac{F_{2c}(\chi)}{\chi^{3/2}(1-\chi)^{1/2}}$ has a pole as $\chi\rightarrow0$ which means that it is necessary to compute the integrals as they are presented in \eqref{b_S phi constraint}. The result is
\begin{equation}
    \frac{\sqrt{N+8}}{2\pi} \left( b_{S\hat{\phi}}^{(1)} + \frac{7N^2+119N+438}{4\sqrt{2N}(N+8)^{3/2}} - \frac{(N+14)\log2}{\sqrt{2N(N+8)}} \right) = 0\,,
\end{equation}
which allows us to read off $b_{S\hat{\phi}}^{(1)}$. Putting everything together, we find
\begin{equation}
    b_{S\hat{\phi}} = \sqrt{\frac{N+8}{2N}} \left( 1 + \varepsilon \left( \frac{(N+14)\log2}{N+8} - \frac{7N^2+119N+438}{4(N+8)^2}\right)\right)\,.
\end{equation}

As for the traceless-symmetric operator $T(u,x)\equiv u^a u^b T_{ab}(x)$, we first consider the case where $T$ is inserted in the bulk, and we have insertions of $\hat{\phi}$ and $\hat{t}$ on the defect.
The correlation function is connected and takes the form $\langle T(u,(0,1)) \hat{\phi}(1) (\hat{u}\cdot\hat{t})(\tau)\rangle_c = 2\frac{(u\cdot\hat{u})(u\cdot\theta)}{1+\tau^2}F(\chi)$, where from \eqref{F_T t phi full}, $F(\chi)$ is given by
\begin{multline}
    F(\chi) = \frac{\sqrt{N+8}}{2\sqrt{2}\pi} \biggl( 1 + \varepsilon \biggl( 2\log2 
        -\frac{5N^2+103N+438}{4(N+8)^2} + \frac{\log\chi}{(N+8)} -\frac{\pi^2-4\arccos^2\sqrt{\chi}}{8\chi}  \biggr) \biggr)\,.
\end{multline}
If we take the two-point function of $T$ with $\hat{\phi}$ to take the form $\langle T(u,(0,1)) \hat{\phi}(0)\rangle_c = (u\cdot\theta)^2 b_{T\hat{\phi}}$, then the identity that we find is
\begin{multline}\label{b_Tphihat calculated}
    -2\frac{C_{\hat{t}}^{-1} b_{T\hat{t}} \lambda_{\hat{t}\hat{t}\hat{\phi}} }{\gamma_{\hat{\phi}}} + \int_0^\frac{1}{2}d\chi \left(\frac{F_{2c}(\chi)}{\chi^{1/2}(1-\chi)^{1/2}} - \frac{C_{\hat{t}}^{-1} b_{T\hat{t}} \lambda_{\hat{t}\hat{t}\hat{\phi}}}{\chi(1-\chi) } \right) + \int_{\frac{1}{2}}^1 d\chi \frac{F_{2c}(\chi)}{\chi^{1/2}(1-\chi)^{1/2}}
    \\ = \sqrt{\frac{N+2}{8}} \left(1+\varepsilon\left(\frac{2(N+7)}{N+8} - \frac{5N^2+103N+438}{4(N+8)^2}\right)\right) = b_{T\hat{\phi}}\,,
\end{multline}
from which we are able to read off the two-point coefficient $b_{T\hat{\phi}}$.

The final three-point function example is $\langle T_{ab} \hat{t}_\imath \hat{t}_\jmath \rangle_c$, the general form of which is given in \eqref{uOtt}. We calculate the correlation function from Feynman diagrams in \eqref{F_T t t full} the find the two channels of the connected correlation function to be given by
\begin{align}
    F_{1c}(\chi) & = \frac{N+8}{2\sqrt{2}\pi^2}\left( 1 + \varepsilon \left( \log2 + \frac{N^2-5N-38}{2(N+8)^2} + \frac{1}{N+8}\log\chi \right) \right)\,,\\
    F_{2c}(\chi) & = -\varepsilon\frac{(N+8)}{16\sqrt{2}\pi^2} (\pi^2-4\arccos^2 \sqrt{\chi})\,.
\end{align}
Working first with the $F_{1c}$ channel, we use \eqref{intF1}. Indeed,
\begin{equation}
    \int_0^1 d\chi \frac{F_{1c}(\chi)}{\chi^{1/2} (1-\chi)^{1/2}} = \frac{N+8}{2\pi\sqrt{2}}\left(1+\varepsilon\left( \frac{N+6}{N+8}\log2 + \frac{N^2-5N-38}{2(N+8)^2}\right)\right) = 2b_{T\hat{t}}\,.
\end{equation}
For the $F_{2c}$ channel, the relevant identity is \eqref{perturbativeF2int}. Performing the calculation, we find
\begin{multline}
    2 \frac{b_{T\hat{t}} \lambda_{\hat{t}\hat{t}\hat{\phi}}}{\gamma_{\hat{\phi}}} + \int_0^\frac{1}{2}d\chi \left(\frac{F_{2c}(\chi)}{\chi^{3/2}(1-\chi)^{1/2}} - \frac{b_{T\hat{\phi}} \lambda_{\hat{t}\hat{t}\hat{\phi}}}{\chi (1-\chi) } \right) + \int_{\frac{1}{2}}^1 d\chi \frac{F_{2c}(\chi)}{\chi^{3/2}(1-\chi)^{1/2}}
    \\ = -\frac{N+8}{2\pi\sqrt{2}}\left(1+\varepsilon\left( \frac{N+6}{N+8}\log2 + \frac{N^2-5N-38}{2(N+8)^2}\right)\right) = -2 b_{T\hat{t}}\,.
\end{multline}
The previous two results could have been used to calculate $b_{T\hat{t}}$ and $b_{T\hat{\phi}}$. Since these quantities were calculated earlier through other integral identities, these two equations serve as self-consistency checks.

\subsection{\texorpdfstring{$\int\langle \cO \hat{\cO} \hat{\cO} \hat{t} \rangle \sim \langle \cO \hat{\cO} \hat{\cO} \rangle$}{m=1, n=1 with two extra defect operators}}

A general bulk--defect--defect--defect correlator has the form
\begin{equation}
    \langle \cO_1(x_1) \hat{\cO}_2(\tau_2) \hat{\cO}_3(\tau_3) \hat{\cO}_4(\tau_4) \rangle = \frac{F_{\cO_1 \hat{\cO}_2 \hat{\cO}_3 \hat{\cO}_4}(u,v) }{ |x_{1\perp}|^{\Delta_1} |\tau_{23}|^{\Delta_2+\Delta_3-\Delta_4} |\tau_{24}|^{\Delta_2-\Delta_3+\Delta_4} |\tau_{34}|^{-\Delta_2+\Delta_3+\Delta_4} }\,,
\end{equation}
where the function $F_{\cO_1 \hat{\cO}_2 \hat{\cO}_3 \hat{\cO}_4}$ now depends on the two cross-ratios
\begin{equation}
    u=\frac{(\tau_{12}^2+x_{1\perp}^2) \tau_{34}^2 }{(\tau_{13}^2+x_{1\perp}^2)\tau_{24}^2}\,,\qquad v = \frac{(\tau_{14}^2+x_{1\perp}^2) \tau_{23}^2 }{(\tau_{13}^2+x_{1\perp}^2)\tau_{24}^2}\,.
\end{equation}
It is also useful to use a complex cross-ratio $z$ that satisfies
\begin{equation}
    z \Bar{z} = u\,,\qquad (1-z)(1-\Bar{z})=v\,.
\end{equation}
The identity that we use is \eqref{m=1n=2SeparatePoints} with the bulk operator $\phi_a$---the connected correlation function $\langle (u\cdot\phi) (\hat{u}_1\cdot\hat{t}) (\hat{u}_2\cdot\hat{t})(\hat{u}_3\cdot\hat{t})\rangle_c$ has contributions from a single diagram---the corresponding Feynman integral is conformally equivalent to the scalar box integral. Therefore, the correlation function takes the form
\begin{equation}\label{phi t t t connected}
    \langle (u\cdot\phi)(0,x_{1\perp}) (\hat{u}_1\cdot\hat{t})(-1) (\hat{u}_2\cdot\hat{t})(1) (\hat{u}_3\cdot\hat{t})(\tau)\rangle_c = -((u\cdot\hat{u}_1)(\hat{u}_2\cdot\hat{u}_3)+\dots)\varepsilon\frac{\sqrt{N+8}z \Bar{z} D(z,\Bar{z})}{8\pi^3(1+x_{1\perp}^2)(1-\tau)^2}\,,
\end{equation}
where the dots denote permutations of the polarization vectors and the scalar box integral is given by
\begin{equation}\label{Bloch-Wigner}
    D(z,\Bar{z}) = \frac{1}{z - \Bar{z}} \left( 2\Li_2(z)-2\Li_2(\Bar{z})+\log z \Bar{z} \log \frac{1-z}{1-\Bar{z}} \right)\,.
\end{equation}
The function in \eqref{phi t t t connected} has no poles in $\tau$ at this order in $\varepsilon$, but rather one must take care of branch cuts when integrating. We proceed by noticing that the configuration used is even in $\tau$, so we therefore restrict the integral to the positive real line. The correlator has branch cuts at $\tau=\pm1$, so we split the integration region into $\tau\in(0,1)$ and $\tau\in(1,\infty)$.
Calculating the primitive and evaluating it in four regimes, $\tau\rightarrow\infty$, $\tau\rightarrow1^+$, $\tau\rightarrow1^-$ and $\tau\rightarrow0$, takes care of the branch cut at $\tau=1$. After making use of the inversion and reflection identities for the di- and trilogarithm, one finds that 
\begin{multline}\label{phi t t t integrated}
    \int_{-\infty}^\infty d\tau  \langle (u\cdot\phi)(0,x_{1\perp}) (\hat{u}_1\cdot\hat{t})(-1) (\hat{u}_2\cdot\hat{t})(1) (\hat{u}_3\cdot\hat{t})(\tau)\rangle_c \\= -((u\cdot\hat{u}_1)(\hat{u}_2\cdot\hat{u}_3)+\dots)\frac{\varepsilon\sqrt{N+8}}{64\pi^2 |x_{1\perp}|} \biggl( \pi^2 - 4\arccos^2 \frac{2|x_{1\perp}|}{1+|x_{1\perp}|^2} \biggr)\,.
\end{multline}
Even though \eqref{phi t t t connected} has no poles in $\tau$ at leading order in $\varepsilon$, there are contact terms at $O(\varepsilon)$ and they need to be included. The contact terms arise from the $\hat{\phi}$ contribution to the $(\hat{u}_1\cdot\hat{t})(-1)\times(\hat{u}_3\cdot\hat{t})(\tau)$ and $(\hat{u}_2\cdot\hat{t})(1)\times(\hat{u}_3\cdot\hat{t})(\tau)$ OPE's. The contribution from the first OPE is given by
\begin{multline}\label{phi t t t contact terms}
    2\frac{C_{\hat{t}} \lambda_{ \hat{t}\hat{t}\hat{\phi} }}{\gamma_{\hat{\phi}}} (\hat{u}_1\cdot\hat{u}_3) \langle (u\cdot\phi)(0,x_{1\perp}) \hat{\phi}(-1) (\hat{u}_2\cdot\hat{t})(1)\rangle_c
    \\ =(u\cdot\hat{u}_2)(\hat{u}_1\cdot\hat{u}_3) \frac{\varepsilon\sqrt{N+8}}{64\pi^2 |x_{1\perp}|}\left( \pi^2 - 4\arccos^2 \frac{2|x_{1\perp}|}{1+|x_{1\perp}|^2} \right)\,,
\end{multline}
with the second contact term obtained by interchanging $\hat{u}_1\leftrightarrow\hat{u}_2$.
Finally, we need the terms from the broken symmetry generator acting on $\phi_a$. We have 
\begin{multline}
    (u\cdot\hat{u}_3)\langle \hat{\phi} (0,x_{1\perp}) (\hat{u}_1\cdot\hat{t})(-1) (\hat{u}_2\cdot\hat{t})(1) \rangle_c - (u\cdot\theta) \langle (\hat{u}_3\cdot\phi)(0,x_{1\perp})(\hat{u}_1\cdot\hat{t})(-1)(\hat{u}_2\cdot\hat{t})(1) \rangle_c \\ = (u\cdot\hat{u}_3)(\hat{u}_1\cdot\hat{u}_2) \frac{\varepsilon\sqrt{N+8}}{64\pi^2 |x_{1\perp}|}\left( \pi^2 - 4\arccos^2 \frac{2|x_{1\perp}|}{1+|x_{1\perp}|^2} \right)\,,
\end{multline}
since the second term vanishes.
Combining this with \eqref{phi t t t integrated} and \eqref{phi t t t contact terms}, we find that the expression vanishes as expected by \eqref{m=1n=2SeparatePoints}.


\section{Conclusion}
Through the breaking of an internal symmetry, the presence of a conformal defect introduces a distinguished set of operators with support on the defect. The tilt operators modify the current conservation equation and lead to powerful relations among connected correlation functions, as well as novel constraints on the CFT data. We develop a framework that allows us to derive relations among CFT data in complete generality by analyzing correlation functions that involve up to a single bulk insertion and up to three defect insertions.

We put our relations to use in two settings. First, the 1/2 BPS Maldacena-Wilson loop in $\cN=4$ SYM. In this setting, our constraints should hold at all values of the 't Hooft coupling and this is exactly what we find with the relation in \eqref{N=4 a~b all couplings}. While other relations should hold for higher point functions, in practice it is more straightforward to work perturbatively at either weak- or strong-coupling. At weak-coupling, we demonstrate the power of our integral identities through self-consistency. With an ansatz of the forms of the three-point function of a 1/2 BPS operator in the bulk and two tilt operators on the defect, we are able to use some existing results to determine CFT data at leading order and next-to-leading order, for example \eqref{bOJphi}.

After making use of our integral identities to highlight discrepancies among existing results at strong coupling, we hope that the tools we have developed will be useful in verifying the self-consistency of existing results, as well as being a valuable way of determining new CFT data in the future through their incorporation in the bootstrap program.

The second setting we study is the magnetic line defect in the $O(N)$ model in $d=4-\varepsilon$ dimensions. Here, we find that we are able to verify a wide array of CFT data that already exists in the literature. Moreover, new CFT data (also in \cite{Girault:2025kzt}) that is calculated using traditional Feynman diagram techniques is found to be compatible with the constraints that we have derived.

Similar integral identities can also be constructed for displacement operators rather than tilts. These follow from the broken conformal generators and, as shown in \cite{upcoming}, often yield even stronger constraints than those derived from tilts; see also \cite{Girault:2025kzt} for related constructions. Anomalies form another key ingredient: they play a crucial role in \cite{upcoming} and are expected to be equally relevant in the present bulk–defect context, although we have not analyzed them here. We leave this to future work.

A further interesting direction arises from a geometric perspective. Tilts are defect exactly marginal operators, permitting deformations along a defect conformal manifold. However, one may encounter additional defect exactly marginal operators beyond tilts, for instance those relating distinct conformal BPS line defects in three dimensions \cite{Drukker:2022ywj, Kong:2022yib, Correa:2019rdk}. It would be interesting to study the integral identities associated with such operators to the curvature of the defect conformal manifold, in analogy with \cite{Drukker:2022pxk}. Moreover, when the bulk theory also contains exactly marginal operators, one can integrate bulk insertions in the presence of the defect and explore the richer structure of the combined bulk–defect conformal manifold.

Beyond symmetry breaking induced by defects, other mechanisms exist, such as in thermal CFTs, where analogous Ward identities are known \cite{Marchetto:2023fcw,Barrat:2024aoa} and lead to similar integral constraints.

\section*{Acknowledgements}
We are grateful to Nadav Drukker and Petr Kravchuk for their collaborative work in~\cite{upcoming}, from which some results are presented here and whose ideas and techniques we have adopted throughout this paper. This project would not have been possible without that work. We also thank them for many inspiring and helpful discussions. We are especially indebted to N.~Drukker for reviewing the draft and providing valuable suggestions. We also would like to acknowledge fruitful discussions with J. Barrat, P. van Vliet and M. F. Paulos.
Z.K. is supported by ERC-2021-CoG---BrokenSymmetries 101044226.

\begin{appendices}

\section{Feynman Integrals and Correlation Functions}\label{Feynman Integrals}

We compute Feynman integrals and two- and three-point functions in the $O(N)$ model in scenarios with a single bulk operator and defect operators with dimensions close to one. The Feynman rules include the free propagator \cite{Cuomo:2021kfm,Gimenez-Grau:2022czc}
\begin{equation}
    \begin{tikzpicture}[scale=0.5,baseline=(vert_cent.base)]
        \draw[dashed,black] (0,1)--(1.5,1);
        \node[inner sep=0pt,outer sep=0pt] (vert_cent) at (0,1) {$\phantom{\cdot}$};
    \end{tikzpicture} \equiv \langle \phi_a(x_1) \phi_b(x_2) \rangle_{\lambda_0 = h_0 = 0} = \frac{\kappa \delta_{ab}}{(x_{12}^2)^{1-\varepsilon/2}}\,,
    \qquad \kappa = \frac{\Gamma(\frac{4-\varepsilon}{2})}{2\pi^{(4-\varepsilon)/2}(2-\varepsilon)}\,,
\end{equation}
and the bulk and defect vertices which respectively are given by
\begin{equation}
    \begin{tikzpicture}[scale=0.5,baseline=(vert_cent.base)]
        \fill[blue] (0,1) circle (3pt);
        \node[inner sep=0pt,outer sep=0pt] (vert_cent) at (0,1) {$\phantom{\cdot}$};
    \end{tikzpicture}\, \equiv -\lambda_0 \int d^d x\,, \qquad
    \begin{tikzpicture}[scale=0.5,baseline=(vert_cent.base)]
        \draw[very thick,black] (-2,1)--(2,1);
        \fill[blue] (-0.1,0.9) rectangle (0.1,1.1);
        \node[inner sep=0pt,outer sep=0pt] (vert_cent) at (0,1) {$\phantom{\cdot}$};
    \end{tikzpicture} \, \equiv -h_0 \int_{-\infty}^\infty d\tau\,,
\end{equation}
where the thick black line denotes the defect. The bulk operators that we consider are $\phi_a$, $S\sim \phi_a\phi_a$ and $T_{ab}\sim\phi_a\phi_b-\tr$, while the defect operators are $\hat{\phi}$ and $\hat{t}_\imath$. The renormalization factors of these operators, along with other integrals that contribute, can be found in the literature, see for example \cite{Cuomo:2021kfm,Gimenez-Grau:2022ebb,Gimenez-Grau:2022czc}.

First, we consider the bulk--defect two-point function that involves a bulk operator of dimension close to two.
In terms of bare fields and Feynman diagrams, this correlation function is given by
\begin{equation}
    \langle (\phi_{0a}\phi_{0b})(x_1) \phi_{0c}(\tau_2) \rangle = 
    \begin{tikzpicture}[scale=0.5,baseline=(vert_cent.base)]
        \draw[very thick,black] (-2,0)--(2,0);
        \draw[dashed,black] (-1,0)--(0,2);
        \draw[dashed,black] (1,0)--(0,2);
        \fill[black] (0,2) circle (3pt);
        \fill[black] (-1,0) circle (3pt);
        \fill[blue] (0.9,-0.1) rectangle (1.1,0.1);
        \node[inner sep=0pt,outer sep=0pt] (vert_cent) at (0,1) {$\phantom{\cdot}$};
    \end{tikzpicture} + 
    \begin{tikzpicture}[scale=0.5,baseline=(vert_cent.base)]
        \draw[very thick,black] (-2,0)--(2,0);
        \draw[dashed,black] (-1,0)--(0,2);
        \draw[dashed,black] (1,0.8)--(0,2);
        \draw[dashed,black] (1,0.8)--(0.5,0);
        \draw[dashed,black] (1,0.8)--(1,0);
        \draw[dashed,black] (1,0.8)--(1.5,0);
        \fill[black] (0,2) circle (3pt);
        \fill[black] (-1,0) circle (3pt);
        \fill[blue] (0.9,-0.1) rectangle (1.1,0.1);
        \fill[blue] (0.4,-0.1) rectangle (0.6,0.1);
        \fill[blue] (1.4,-0.1) rectangle (1.6,0.1);
        \fill[blue] (1,0.8) circle (3pt);
        \node[inner sep=0pt,outer sep=0pt] (vert_cent) at (0,1) {$\phantom{\cdot}$};
    \end{tikzpicture} + 
    \begin{tikzpicture}[scale=0.5,baseline=(vert_cent.base)]
        \draw[very thick,black] (-2,0)--(2,0);
        \draw[dashed,black] (-1,0)--(0,2);
        \draw[dashed,black] (1,0.8)--(0,2);
        \draw[dashed,black] (1,0.8)--(0.5,0);
        \draw[dashed,black] (1,0.8)--(1,0);
        \draw[dashed,black] (1,0.8)--(1.5,0);
        \fill[black] (0,2) circle (3pt);
        \fill[black] (1,0) circle (3pt);
        \fill[blue] (-1.1,-0.1) rectangle (-0.9,0.1);
        \fill[blue] (0.4,-0.1) rectangle (0.6,0.1);
        \fill[blue] (1.4,-0.1) rectangle (1.6,0.1);
        \fill[blue] (1,0.8) circle (3pt);
        \node[inner sep=0pt,outer sep=0pt] (vert_cent) at (0,1) {$\phantom{\cdot}$};
    \end{tikzpicture} +
    \begin{tikzpicture}[scale=0.5,baseline=(vert_cent.base)]
        \draw[very thick,black] (-2,0)--(2,0);
        \draw[dashed,black] (-1,0)--(0,1);
        \draw[dashed,black] (1,0)--(0,1);
        \draw[dashed] (0,1) to[out=-45,in=45](0,2);
        \draw[dashed] (0,1) to[out=-135,in=135](0,2);
        \fill[blue] (0,1) circle (3pt);
        \fill[black] (0,2) circle (3pt);
        \fill[black] (-1,0) circle (3pt);
        \fill[blue] (0.9,-0.1) rectangle (1.1,0.1);
        \node[inner sep=0pt,outer sep=0pt] (vert_cent) at (0,1) {$\phantom{\cdot}$};
    \end{tikzpicture}\,.
\end{equation}
Expressions for the subdiagrams in the first three terms can be found in \cite{Gimenez-Grau:2022ebb}. The fourth diagram is new and is given by
\begin{equation}
    \begin{tikzpicture}[scale=0.7,baseline=(vert_cent.base)]
        \draw[very thick,black] (-2,0)--(2,0);
        \draw[dashed,black] (-1,0)--(0,1);
        \draw[dashed,black] (1,0)--(0,1);
        \draw[dashed] (0,1) to[out=-45,in=45](0,2);
        \draw[dashed] (0,1) to[out=-135,in=135](0,2);
        \fill[blue] (0,1) circle (3pt);
        \fill[black] (0,2) circle (3pt);
        \fill[black] (-1,0) circle (3pt);
        \fill[blue] (0.9,-0.1) rectangle (1.1,0.1);
        \node[inner sep=0pt,outer sep=0pt] (vert_cent) at (0,1) {$\phantom{\cdot}$};
    \end{tikzpicture}
    \propto \int \frac{d\tau_3 d^d x_4}{( \left(x_{14}^2\right)^2 x_{24}^2 x_{34}^2 )^{\Delta_\phi}}\,,
\end{equation}
where black circles denote external operators. We take $x_{1,4} = (\tau_{1,4}, x_{1,4\perp})$ and $x_{2,3} = (\tau_{2,3},0)$. Also, $\Delta_\phi = 1-\frac{\varepsilon}{2}$ since these integrals are computed for bare operators. This integral is divergent in $d=4$ and must therefore be computed in $d=4-\varepsilon$. The diagram is $O(\lambda)$, which means that it is enough to calculate the integral to $O(\varepsilon^0)$.
To this end, we introduce Schwinger parameters $u_1$, $u_2$ and $u_3$. It is then trivial to carry out the integrals over $\tau_3$ and $x_4$. One of the Schwinger parameters, say $u_3$, can be integrated by first introducing a resolution of the identity, $1=\int_0^\infty dq \,\delta(q-u_3)$, and then rescaling the Schwinger parameters as $u_i \rightarrow q u_i$. The parameter $q$ is easy to integrate away and the remaining delta function $\delta(1-u_3)$ allows the $u_3$ integral to be done trivially. 
To progress, we need to identify where the integral is divergent using the procedure of analytic regularization as outlined in \cite{Panzer:2014gra,Panzer:2015ida}. The only divergence is when $u_1\rightarrow\infty$. The prescription to deal with this is to introduce another delta function, $1=\int_0^\infty dq \, \delta(q-u_1^{-1})$, rescale $u_1 \rightarrow q^{-1}u_1$ and then partially integrate with respect to $q$. It is straightforward to check that the surface terms vanish and what remains is a factor proportional to $\frac{1}{\varepsilon}$ and a convergent integral that can be expanded in $\varepsilon$ and evaluated term-by-term. The final result is
\begin{equation}
    \int \frac{d\tau_3 d^d x_4}{( \left(x_{14}^2\right)^2 x_{24}^2 x_{34}^2 )^{\Delta_\phi}} = \frac{\pi^3}{|x_{1\perp}|^{1-2\varepsilon} (x_{12}^2)^{1-\frac{\varepsilon}{2}} } \left( \frac{2}{\varepsilon} +2+4\log2-\gamma_E-\log\pi\right)\,.
\end{equation}
This allows us to compute $b_{T \hat{t}}$ and $b_{T \hat{\phi}}$, which are given in \eqref{b_Tt} and \eqref{b_Tphihat calculated} respectively.

Next, we calculate the bulk--defect--defect three-point function where all operators have dimension close to one. The bare correlation function is given by
\begin{equation}
    \langle \phi_{0a}(x_0)\phi_{0b}(\tau_1) \phi_{0c}(\tau_2) \rangle = 
    \begin{tikzpicture}[scale=0.5,baseline=(vert_cent.base)]
        \draw[very thick,black] (-2,0)--(2,0);
        \draw[dashed,black] (-1,0)--(-1,2);
        \draw[dashed] (0,0) to[out=60,in=120](1,0);
        \fill[black] (-1,2) circle (3pt);
        \fill[black] (0,0) circle (3pt);
        \fill[black] (1,0) circle (3pt);
        \fill[blue] (-1.1,-0.1) rectangle (-0.9,0.1);
        \node[inner sep=0pt,outer sep=0pt] (vert_cent) at (0.5,1) {$\phantom{\cdot}$};
    \end{tikzpicture} + 
    \begin{tikzpicture}[scale=0.5,baseline=(vert_cent.base)]
        \draw[very thick,black] (-2,0)--(2,0);
        \draw[dashed,black] (-1,0.8)--(-1,2);
        \draw[dashed] (0.5,0) to[out=60,in=120](1.5,0);
        \draw[dashed,black] (-1,0.8)--(-0.5,0);
        \draw[dashed,black] (-1,0.8)--(-1,0);
        \draw[dashed,black] (-1,0.8)--(-1.5,0);
        \fill[black] (-1,2) circle (3pt);
        \fill[black] (0.5,0) circle (3pt);
        \fill[black] (1.5,0) circle (3pt);
        \fill[blue] (-1.1,-0.1) rectangle (-0.9,0.1);
        \fill[blue] (-1.6,-0.1) rectangle (-1.4,0.1);
        \fill[blue] (-0.6,-0.1) rectangle (-0.4,0.1);
        \fill[blue] (-1,0.8) circle (3pt);
        \node[inner sep=0pt,outer sep=0pt] (vert_cent) at (0,1) {$\phantom{\cdot}$};
    \end{tikzpicture} + 
    \begin{tikzpicture}[scale=0.5,baseline=(vert_cent.base)]
        \draw[very thick,black] (-2,0)--(2,0);
        \draw[dashed,black] (-1,0)--(-1,2);
        \draw[dashed] (0.75,0.6)--(1.5,0);
        \draw[dashed] (0,0)--(0.75,0.6);
        \draw[dashed] (0.75,0.6)--(0.5,0);
        \draw[dashed] (0.75,0.6)--(1,0);
        \fill[black] (-1,2) circle (3pt);
        \fill[black] (0,0) circle (3pt);
        \fill[black] (1.5,0) circle (3pt);
        \fill[blue] (0.75,0.6) circle (3pt);
        \fill[blue] (-1.1,-0.1) rectangle (-0.9,0.1);
        \fill[blue] (0.4,-0.1) rectangle (0.6,0.1);
        \fill[blue] (0.9,-0.1) rectangle (1.1,0.1);
        \node[inner sep=0pt,outer sep=0pt] (vert_cent) at (0.5,1) {$\phantom{\cdot}$};
    \end{tikzpicture} +
    \begin{tikzpicture}[scale=0.5,baseline=(vert_cent.base)]
        \draw[very thick,black] (-2,0)--(2,0);
        \draw[dashed,black] (-1,0)--(0,1);
        \draw[dashed,black] (1,0)--(0,1);
        \draw[dashed] (0,1)--(0,2);
        \draw[dashed] (0,0)--(0,1);
        \fill[blue] (0,1) circle (3pt);
        \fill[black] (0,2) circle (3pt);
        \fill[black] (-1,0) circle (3pt);
        \fill[black] (1,0) circle (3pt);
        \fill[blue] (-0.1,-0.1) rectangle (0.1,0.1);
        \node[inner sep=0pt,outer sep=0pt] (vert_cent) at (0,1) {$\phantom{\cdot}$};
    \end{tikzpicture}\,,
\end{equation}
where we place the bulk operator at $x_{0} = (\tau_{0}, x_{0\perp})$ and the defect operators at $x_{1,2} = (\tau_{1,2},0)$. In this setup, there is a single cross-ratio given by $\chi$, defined in \eqref{chi cross ratio}.
The fourth diagram is new and is finite in $d=4$. It is given by
\begin{equation}
    \begin{tikzpicture}[scale=0.7,baseline=(vert_cent.base)]
        \draw[very thick,black] (-2,0)--(2,0);
        \draw[dashed,black] (-1,0)--(0,1);
        \draw[dashed,black] (0,0)--(0,2);
        \draw[dashed,black] (1,0)--(0,1);
        \fill[blue] (-0.1,-0.1) rectangle (0.1,0.1);
        \fill[blue] (0,1) circle (3pt);
        \fill[black] (0,2) circle (3pt);
        \fill[black] (-1,0) circle (3pt);
        \fill[black] (1,0) circle (3pt);
        \node[inner sep=0pt,outer sep=0pt] (vert_cent) at (0,1) {$\phantom{\cdot}$};
    \end{tikzpicture}
    \propto \int \frac{d \tau_3 d^4 x_4}{x_{04}^2 x_{14}^2 x_{24}^2 x_{34}^2} = \frac{\pi^3}{2} \frac{\cI_{1,2}(\chi)}{|x_{0\perp}| \tau_{12}^2} \,,
\end{equation}
where $x_3=(\tau_3,0)$ and $x_4=(\tau_4,x_{4\perp})$ are the points being integrated over. We proceed to evaluate this integral by first taking $\tau_1 \rightarrow \infty$, and computing
\begin{equation}
    \int \frac{ d\tau_3 d^4 x_4}{x_{04}^2 x_{24}^2 x_{34}^2 } = \pi^3 \frac{\pi^2 - 4 \arctan^2 \left( \frac{\tau_{02}}{|x_{0\perp}|} \right) }{2|x_{0\perp}|}\,.
\end{equation}
To evaluate this integral, one first introduces Schwinger parameters $u_1, u_2, u_3$.
It is then straightforward to integrate over $\tau_3$, $\tau_4$ and $x_{4\perp}$.
To integrate over the Schwinger parameters, it is useful to turn the integral into a projective integral by introducing $1=\int_0^\infty dq \, \delta(q-u_1-u_2-u_3)$ and rescaling each $u_i \rightarrow q u_i$. It is elementary to now integrate over the auxiliary variable $q$, while the delta function makes one of the Schwinger parameter integrals, say $u_3$, trivial. The final integrals over $0<u_{1,2}<1$ subject to $u_1+u_2<1$ are less straightforward, but can be calculated in Mathematica.
We are then able to reconstruct the original integral and conclude
\begin{equation}
    \mathcal{I}_{1,2}(\chi) = \pi^2 - 4 \arccos^2 \left( \sqrt{\chi} \right)\,.
\end{equation}
When combining this result with the disconnected diagrams, which have been previously calculated in \cite{Gimenez-Grau:2022ebb,Gimenez-Grau:2022czc}, we find the following correlators,
\begin{align}
    \label{F_phi t t full}F_{\phi \hat{t} \hat{t}} & = (u\cdot\theta)(\hat{u}_1\cdot\hat{u}_2) \frac{(N+8)^{\frac{3}{2}}}{8\pi^2\chi}\left( 1 + \varepsilon \left( \frac{7N^2+55N+190}{4(N+8)^2} + \frac{\log2}{2} - \frac{\cI_{1,2}(\chi)}{2(N+8)} \right) \right)\,,\\
    \label{F_phi phi t full}F_{\phi \hat{\phi} \hat{t} } & = -(u\cdot\hat{u}_1)\frac{\cI_{1,2}(\chi)}{8\pi\chi}\varepsilon\,,\\
    \label{F_phi phi phi full}F_{\phi \hat{\phi} \hat{\phi}} & = (u\cdot\theta) \frac{\sqrt{N+8}}{2\chi} \left( 1 + \varepsilon\left( \frac{N^2-3N-22}{4(N+8)^2} - \log\left( \frac{\chi}{\sqrt{2}} \right) - \frac{3\cI_{1,2}(\chi)}{2(N+8)} \right) \right)\,,
\end{align}
where we have contracted the bulk operator $\phi_a$ with a polarization vector $u$ and the tilt operators with polarization vectors $\hat{u}_n$.

Finally, we consider the case where the dimension of the bulk operator is close to two. In terms of Feynman diagrams, we have
\begin{multline}
    \langle (\phi_{0a}\phi_{0b})(x_0)\phi_{0c}(\tau_1) \phi_{0d}(\tau_2) \rangle = 
    \begin{tikzpicture}[scale=0.5,baseline=(vert_cent.base)]
        \draw[very thick,black] (-2,0)--(2,0);
        \draw[dashed,black] (-1.25,0)--(-0.75,2);
        \draw[dashed,black] (-0.25,0)--(-0.75,2);
        \fill[blue] (-1.35,-0.1) rectangle (-1.15,0.1);
        \fill[blue] (-0.35,-0.1) rectangle (-0.15,0.1);
        \fill[black] (-0.75,2) circle (3pt);
        \fill[black] (0.5,0) circle (3pt);
        \fill[black] (1.5,0) circle (3pt);
        \draw[dashed] (0.5,0) to[out=60,in=120](1.5,0);
        \node[inner sep=0pt,outer sep=0pt] (vert_cent) at (0,1) {$\phantom{\cdot}$};
    \end{tikzpicture} + 
    \begin{tikzpicture}[scale=0.5,baseline=(vert_cent.base)]
        \draw[very thick,black] (-2,0)--(2,0);
        \draw[dashed,black] (-1,0.8)--(-1,2);
        \draw[dashed,black] (-1,2)--(0,0);
        \draw[dashed] (0.5,0) to[out=60,in=120](1.5,0);
        \draw[dashed,black] (-1,0.8)--(-0.5,0);
        \draw[dashed,black] (-1,0.8)--(-1,0);
        \draw[dashed,black] (-1,0.8)--(-1.5,0);
        \fill[black] (-1,2) circle (3pt);
        \fill[blue] (-0.1,-0.1) rectangle (0.1,0.1);
        \fill[black] (0.5,0) circle (3pt);
        \fill[black] (1.5,0) circle (3pt);
        \fill[blue] (-1.1,-0.1) rectangle (-0.9,0.1);
        \fill[blue] (-1.6,-0.1) rectangle (-1.4,0.1);
        \fill[blue] (-0.6,-0.1) rectangle (-0.4,0.1);
        \fill[blue] (-1,0.8) circle (3pt);
        \node[inner sep=0pt,outer sep=0pt] (vert_cent) at (0,1) {$\phantom{\cdot}$};
    \end{tikzpicture} +
    \begin{tikzpicture}[scale=0.5,baseline=(vert_cent.base)]
        \draw[very thick,black] (-2,0)--(2,0);
        \draw[dashed,black] (-1.25,0)--(-0.75,2);
        \draw[dashed,black] (-0.25,0)--(-0.75,2);
        \fill[blue] (-1.35,-0.1) rectangle (-1.15,0.1);
        \fill[blue] (-0.35,-0.1) rectangle (-0.15,0.1);
        \fill[black] (-0.75,2) circle (3pt);
        \fill[black] (0.25,0) circle (3pt);
        \fill[black] (1.75,0) circle (3pt);
        \fill[blue] (1,0.6) circle (3pt);
        \draw[dashed] (1,0.6)--(1.75,0);
        \draw[dashed] (0.25,0)--(1,0.6);
        \draw[dashed] (1,0.6)--(0.75,0);
        \draw[dashed] (1,0.6)--(1.25,0);
        \fill[blue] (0.65,-0.1) rectangle (0.85,0.1);
        \fill[blue] (1.15,-0.1) rectangle (1.35,0.1);
        \node[inner sep=0pt,outer sep=0pt] (vert_cent) at (0,1) {$\phantom{\cdot}$};
    \end{tikzpicture} + 
    \begin{tikzpicture}[scale=0.5,baseline=(vert_cent.base)]
        \draw[very thick,black] (-2,0)--(2,0);
        \draw[dashed,black] (-1.15,0)--(-0.75,1);
        \draw[dashed,black] (-0.35,0)--(-0.75,1);
        \draw[dashed] (-0.75,1) to[out=-45,in=45](-0.75,2);
        \draw[dashed] (-0.75,1) to[out=-135,in=135](-0.75,2);
        \fill[blue] (-0.75,1) circle (3pt);
        \fill[black] (-0.75,2) circle (3pt);
        \fill[blue] (-1.25,-0.1) rectangle (-1.05,0.1);
        \fill[blue] (-0.45,-0.1) rectangle (-0.25,0.1);
        \fill[black] (0.5,0) circle (3pt);
        \fill[black] (1.5,0) circle (3pt);
        \draw[dashed] (0.5,0) to[out=60,in=120](1.5,0);
        \node[inner sep=0pt,outer sep=0pt] (vert_cent) at (0,1) {$\phantom{\cdot}$};
    \end{tikzpicture}
    \\ + \begin{tikzpicture}[scale=0.5,baseline=(vert_cent.base)]
        \draw[very thick,black] (-2,0)--(2,0);
        \draw[dashed,black] (-1,0)--(0,2);
        \draw[dashed,black] (1,0)--(0,2);
        \fill[black] (0,2) circle (3pt);
        \fill[black] (-1,0) circle (3pt);
        \fill[black] (1,0) circle (3pt);
        \node[inner sep=0pt,outer sep=0pt] (vert_cent) at (0,1) {$\phantom{\cdot}$};
    \end{tikzpicture} + 
    \begin{tikzpicture}[scale=0.5,baseline=(vert_cent.base)]
        \draw[very thick,black] (-2,0)--(2,0);
        \draw[dashed,black] (-0.5,0.8)--(-0.5,2);
        \draw[dashed,black] (-0.5,2)--(1,0);
        \draw[dashed,black] (-0.5,0.8)--(-1,0);
        \draw[dashed,black] (-0.5,0.8)--(-0.5,0);
        \draw[dashed,black] (-0.5,0.8)--(-0,0);
        \fill[black] (-0.5,2) circle (3pt);
        \fill[blue] (0.9,-0.1) rectangle (1.1,0.1);
        \fill[blue] (-0.6,-0.1) rectangle (-0.4,0.1);
        \fill[blue] (-0.5,0.8) circle (3pt);
        \fill[black] (-1,0) circle (3pt);
        \fill[black] (0,0) circle (3pt);
        \node[inner sep=0pt,outer sep=0pt] (vert_cent) at (0,1) {$\phantom{\cdot}$};
    \end{tikzpicture} + 
    \begin{tikzpicture}[scale=0.5,baseline=(vert_cent.base)]
        \draw[very thick,black] (-2,0)--(2,0);
        \draw[dashed,black] (-0.5,0.8)--(-0.5,2);
        \draw[dashed,black] (-0.5,2)--(1,0);
        \draw[dashed,black] (-0.5,0.8)--(-1,0);
        \draw[dashed,black] (-0.5,0.8)--(-0.5,0);
        \draw[dashed,black] (-0.5,0.8)--(-0,0);
        \fill[black] (-0.5,2) circle (3pt);
        \fill[blue] (-0.1,-0.1) rectangle (0.1,0.1);
        \fill[blue] (-1.1,-0.1) rectangle (-0.9,0.1);
        \fill[blue] (-0.5,0.8) circle (3pt);
        \fill[black] (-0.5,0) circle (3pt);
        \fill[black] (1,0) circle (3pt);
        \node[inner sep=0pt,outer sep=0pt] (vert_cent) at (0,1) {$\phantom{\cdot}$};
    \end{tikzpicture} + 
    \begin{tikzpicture}[scale=0.5,baseline=(vert_cent.base)]
        \draw[very thick,black] (-2,0)--(2,0);
        \draw[dashed,black] (-1,0)--(0,1);
        \draw[dashed,black] (1,0)--(0,1);
        \draw[dashed] (0,1) to[out=-45,in=45](0,2);
        \draw[dashed] (0,1) to[out=-135,in=135](0,2);
        \fill[blue] (0,1) circle (3pt);
        \fill[black] (0,2) circle (3pt);
        \fill[black] (-1,0) circle (3pt);
        \fill[black] (1,0) circle (3pt);
        \node[inner sep=0pt,outer sep=0pt] (vert_cent) at (0,1) {$\phantom{\cdot}$};
    \end{tikzpicture}\,,
\end{multline}
where in the first line we have the disconnected diagrams and in the second line we have the connected diagrams.
The final diagram is the only new one that we need to calculate and it is given by
\begin{equation}
    \begin{tikzpicture}[scale=0.7,baseline=(vert_cent.base)]
        \draw[very thick,black] (-2,0)--(2,0);
        \draw[dashed,black] (-1,0)--(0,1);
        \draw[dashed,black] (1,0)--(0,1);
        \draw[dashed] (0,1) to[out=-45,in=45](0,2);
        \draw[dashed] (0,1) to[out=-135,in=135](0,2);
        \fill[blue] (0,1) circle (3pt);
        \fill[black] (0,2) circle (3pt);
        \fill[black] (-1,0) circle (3pt);
        \fill[black] (1,0) circle (3pt);
        \node[inner sep=0pt,outer sep=0pt] (vert_cent) at (0,1) {$\phantom{\cdot}$};
    \end{tikzpicture}
    \propto \int \frac{d^d x_3}{( \left(x_{03}^2\right)^2 x_{13}^2 x_{23}^2 )^{\Delta_\phi}}\,,
\end{equation}
where $x_3 = (\tau_3,x_{3\perp})$ is the point being integrated over.
This integral is also divergent in $d=4$ and must be computed in $d=4-\varepsilon$. Since the diagram is $O (\lambda)$, it suffices to compute this integral to $O (\varepsilon^0)$. 
After we introduce the three Schwinger parameters $u_{1,2,3}$, it is straightforward to integrate $\tau_3$ along the line and $x_{3\perp}$ over the $(d-1)$-dimensional bulk. 
To proceed, we introduce $1=\int_0^\infty dq \delta(q-u_3)$ and then rescale $u_i \rightarrow q u_i$. 
The integral over $q$ is easy and the delta function allows us to trivially integrate over $u_3$. 
Once again, to progress we need to determine where the integral is divergent and make use of analytic regularization.
As before, the only place where the integral is divergent is the contribution from $u_1\rightarrow\infty$. 
We therefore deal with this by introducing another delta function, $1=\int_0^\infty dq\, \delta \left(q-u_1^{-1}\right)$, rescaling $u_1 \rightarrow u_1 q^{-1}$ and partially integrating with respect to $q$. 
After determining that the surface terms vanish, we are able to expand the now convergent integral and evaluate term-by-term. The final result is
\begin{equation}
    \int \frac{d^d x_4}{( \left(x_{14}^2\right)^2 x_{24}^2 x_{34}^2 )^{\Delta_\phi}} = \frac{\pi^2}{x_{12}^{2-2\varepsilon} x_{13}^{2-2\varepsilon} t_{23}^\varepsilon} \left( \frac{2}{\varepsilon} + 2 - \gamma_E - \log\pi \right)\,.
\end{equation}
We now report all non-zero correlation functions involving insertions of either $S$ or $T_{ab}$ in the bulk with two defect insertions with operators of dimension close to one,
\begin{align}
\begin{split}
F_{S \hat{\phi} \hat{\phi}} & = \sqrt{\frac{2}{N}} \biggl( 1 + \frac{N+8}{8\chi} - \varepsilon \biggl( \frac{7N+50}{2(N+8)} + \frac{13N+38}{16(N+8)\chi} - 3 \frac{1+4\chi}{4\chi} \log2 
\\ & \qquad + \left(\frac{N+8}{8\chi} - \frac{N+2}{2(N+8)} \right) \log\chi + \frac{3\cI_{1,2}(\chi)}{8\chi} \biggr) \biggr)\,,
\end{split}\\
\begin{split}
\label{F_S t t full}F_{S \hat{t}_\imath \hat{t}_\jmath} &= (\hat{u}_1\cdot\hat{u}_2)\frac{N+8}{2\pi^2\sqrt{2N}} \biggl( 1 + \frac{N+8}{8\chi} - \varepsilon \biggl( \frac{13N+38}{2(N+8)^2} - \frac{3N^2+16N+68}{16(N+8)\chi} 
\\ & \qquad - \left( 1 + \frac{3}{4\chi} \right) \log2 - \frac{N+2}{2(N+8)}\log\chi + \frac{\cI_{1,2}(\chi)}{8\chi} \biggr) \biggr)\,,
\end{split}\\
\begin{split}
F_{T \hat{\phi} \hat{\phi} } &= (u\cdot\theta)^2\biggl( 1 + \frac{N+8}{8\chi} - \varepsilon \biggl( \frac{3N+25}{N+8} - \frac{N^2-5N-38}{16(N+8)\chi} 
\\ & \qquad - \left(3+\frac{N+6}{8\chi}\right)\log2 + \left(\frac{N+8}{8\chi}-\frac{1}{N+8}\right)\log\chi + \frac{3\cI_{1,2}(\chi)}{8\chi} \biggr) \biggr)\,,
\end{split}\\
\begin{split}
\label{F_T t phi full}F_{T \hat{t} \hat{\phi}} &= (u\cdot\hat{u}_1)(u\cdot\theta)\frac{\sqrt{N+8}}{\pi\sqrt{2}} \biggl( 1-\varepsilon\biggl( \frac{5N^2+103N+438}{4(N+8)^2} -2\log2 \\& \qquad -\frac{1}{N+8}\log\chi + \frac{\cI_{1,2}(\chi)}{8\chi} \biggr) \biggr)\,,
\end{split}\\
\label{F_T t t full}F_{T \hat{t}\hat{t}} &= (u\cdot\hat{u}_1)(u\cdot\hat{u}_2) \frac{N+8}{2\pi^2\sqrt{2}} \left( 1 + \varepsilon \left( \frac{N^2-5N-38}{2(N+8)^2} + \log2+\frac{1}{N+8}\log\chi \right) \right) + \\ & \qquad (u\cdot\theta)^2(\hat{u}_1\cdot\hat{u}_2) \frac{(N+8)^2}{16\sqrt{2}\pi^2\chi} \left( 1 +\varepsilon \left( \frac{2N^2+12N+34}{(N+8)^2} +\frac{N+6}{N+8}\log2 -\frac{\cI_{1,2}(\chi)}{N+8} \right) \right)\,,
\end{align}
where we have again contracted the expressions with null polarization vectors.

\section{Connected Correlators}
\label{sec:connected}
As stated in Section~\ref{sec:introduction}, the definition of the connected correlator is given in \eqref{connectedbulkdefect},
\begin{multline}
\label{connectedbulkdefectrepeat}
\langle \cO_{\alpha_1} (x_1) \cdots \cO_{\alpha_m} (x_m) \hat{t}_{i_1} (\tau_1) \cdots \hat{t}_{i_n} (\tau_n) \rangle_c \\
=\frac{\delta}{\delta r^{\alpha_1} (x_1)} \cdots \frac{\delta}{\delta r^{\alpha_m} (x_m)} \frac{\delta}{\delta w^{i_1}(\tau_1)} \cdots \frac{\delta}{\delta w^{i_n}(\tau_n)} \log Z[r,w]\Big|_{r=w=0}\,.
\end{multline}
Consider the set of operators
\bal
S=\{\cO_{\alpha_1}(x_1), \cdots, \cO_{\alpha_m}(x_m), \hat{\cO}_{i_1}(\tau_1),\cdots, \hat{\cO}_{i_n}(\tau_n)\}
\eal
which collects all bulk and defect operators appearing in the correlators---we primarily consider $\hat{\cO}=\hat{t}$, while in Section \ref{sec:O(N)} we also have $\hat{\cO}=\hat{\phi}$. With this notation, \eqref{connectedbulkdefectrepeat} may be rewritten as
\bal
{}&\langle \cO_{\alpha_1} (x_1) \cdots \cO_{\alpha_m} (x_m) \hat{\cO}_{i_1} (\tau_1) \cdots \hat{\cO}_{i_n} (\tau_n) \rangle_c =\sum_k \sum_{\substack{S_1 \cup \cdots \cup S_k =S \\ S_i \cap S_j =\varnothing}} (-1)^{k-1} (k-1)! \langle S_1 \rangle \cdots \langle S_k \rangle\,,
\eal
where $S_i$ denotes a proper subset of $S$. We now analyze several simple cases as illustrative examples.
\subsection{\texorpdfstring{$\boldsymbol{m=0}$}{m=0}}
When there are no bulk operator insertions, we consider defect two-, three- and four-point functions. Since defect one-point functions vanish, the case of the two-point function is trivial,
\bal
\langle \hat{\cO}_{i_1}(\tau_1) \hat{\cO}_{i_2} (\tau_2) \rangle_c =\langle \hat{\cO}_{i_1}(\tau_1) \hat{\cO}_{i_2} (\tau_2) \rangle -\langle \hat{\cO}_{i_1}(\tau_1) \rangle \langle \hat{\cO}_{i_2} (\tau_2) \rangle =\langle \hat{\cO}_{i_1}(\tau_1) \hat{\cO}_{i_2} (\tau_2) \rangle\,.
\eal
We see that the full defect two-point function is connected. For the same reason, the full three-point function is also connected,
\bal
\langle \hat{\cO}_{i_1}(\tau_1) \hat{\cO}_{i_2} (\tau_2) \hat{\cO}_{i_3} (\tau_3) \rangle_c = \langle \hat{\cO}_{i_1}(\tau_1) \hat{\cO}_{i_2} (\tau_2) \hat{\cO}_{i_3} (\tau_3)\rangle\,.
\eal
As for the four-point function, the disconnected components are the three products of two-point functions that can be constructed from pairs of operators that are being considered,
\begin{multline}
\langle \hat{\cO}_{i_1}(\tau_1) \hat{\cO}_{i_2} (\tau_2) \hat{\cO}_{i_3}(\tau_3) \hat{\cO}_{i_4} (\tau_4) \rangle_c \\
=\langle \hat{\cO}_{i_1}(\tau_1) \hat{\cO}_{i_2} (\tau_2) \hat{\cO}_{i_3}(\tau_3) \hat{\cO}_{i_4} (\tau_4) \rangle -\langle \hat{\cO}_{i_1}(\tau_1) \hat{\cO}_{i_2} (\tau_2) \rangle \langle \hat{\cO}_{i_3}(\tau_3) \hat{\cO}_{i_4} (\tau_4) \rangle \\
-\langle \hat{\cO}_{i_1}(\tau_1) \hat{\cO}_{i_3}(\tau_3) \rangle \langle \hat{\cO}_{i_2} (\tau_2) \hat{\cO}_{i_4} (\tau_4) \rangle -\langle \hat{\cO}_{i_1}(\tau_1) \hat{\cO}_{i_4} (\tau_4) \rangle \langle \hat{\cO}_{i_2} (\tau_2) \hat{\cO}_{i_3}(\tau_3)  \rangle\,.
\end{multline}
It can be that some of these two-point functions are zero due to symmetry requirements.

\subsection{\texorpdfstring{$\boldsymbol{m=1}$}{m=1}}
In the presence of the defect, bulk operators admit a non-zero one-point function which is naturally connected,
\bal
\langle \cO_{I_1} (x_1) \rangle_c &=\langle \cO_{I_1} (x_1) \rangle\,.
\eal
Once again, thanks to vanishing defect one-point functions, the bulk--defect two-point function is connected,
\bal
\langle \cO_{I_1} (x_1) \hat{\cO}_{i_1}(\tau_1) \rangle_c &=\langle \cO_{I_1} (x_1) \hat{\cO}_{i_1}(\tau_1) \rangle\,.
\eal
For the case with two defect insertions, the only non-zero disconnected contribution can come from the product of the bulk one-point function with the defect two-point function, 
\bal
\langle \cO_{I_1} (x_1) \hat{\cO}_{i_1}(\tau_1) \hat{\cO}_{i_2}(\tau_2) \rangle_c &=\langle \cO_{I_1} (x_1) \hat{\cO}_{i_1}(\tau_1) \hat{\cO}_{i_2}(\tau_2)\rangle -  \langle \cO_{I_1} (x_1) \rangle \langle \hat{\cO}_{i_1}(\tau_1) \hat{\cO}_{i_2}(\tau_2)\rangle\,.
\eal
Finally, when we have three operators on the defect there are two types of disconnected contributions. The first is the product of the bulk one-point function with the defect three-point function, and the other is the product of a bulk--defect two-point function with a defect two-point function. Since there are three defect operators, the latter contributes three terms,
\begin{multline}
\langle \cO_{I_1} (x_1) \hat{\cO}_{i_1}(\tau_1) \hat{\cO}_{i_2}(\tau_2) \hat{\cO}_{i_3}(\tau_3) \rangle_c =\langle \cO_{I_1} (x_1) \hat{\cO}_{i_1}(\tau_1) \hat{\cO}_{i_2}(\tau_2) \hat{\cO}_{i_3}(\tau_3) \rangle \\
- \langle \cO_{I_1} (x_1) \rangle \langle \hat{\cO}_{i_1}(\tau_1) \hat{\cO}_{i_2}(\tau_2) \hat{\cO}_{i_3}(\tau_3)\rangle
-\langle \cO_{I_1} (x_1) \hat{\cO}_{i_1}(\tau_1) \rangle \langle \hat{\cO}_{i_2}(\tau_2) \hat{\cO}_{i_3}(\tau_3) \rangle \\
-\langle \cO_{I_1} (x_1) \hat{\cO}_{i_2}(\tau_2) \rangle \langle \hat{\cO}_{i_1}(\tau_1) \hat{\cO}_{i_3}(\tau_3) \rangle -\langle \cO_{I_1} (x_1) \hat{\cO}_{i_3}(\tau_3) \rangle \langle \hat{\cO}_{i_1}(\tau_1) \hat{\cO}_{i_2}(\tau_2) \rangle\,.
\end{multline}

\end{appendices}

\bibliographystyle{utphys2}
\bibliography{refs}
\end{document}